\documentclass[seceq]{ptptex}

\usepackage{graphicx}
\usepackage{wrapft}

\font\mbf=cmmib10  %%%scaled \magstep 1
\font\calg=cmsy10  %%%scaled \magstep 1
\def\RMNa{\uppercase\expandafter{\romannumeral 1}} 
\def\RMNb{\uppercase\expandafter{\romannumeral 2}}
\def\RMNc{\uppercase\expandafter{\romannumeral 3}}
\def\RMNd{\uppercase\expandafter{\romannumeral 4}}
\title{%        %You can use \\ for explicit line-break.
Resonances in $^{28}$Si$+^{28}$Si. {\uppercase\expandafter{\romannumeral 2}}
}

\subtitle{Analyses for the Angular Distributions and Angular 
Correlations}    %Use this when you want a subtitle.

\author{%       %Use \scshape for the family name.
Eiji \textsc{Uegaki}$^1$%
 and Yasuhisa \textsc{Abe}$^2$
} 

\inst{%     %Affiliation, neglected when [addenda] or [errata].
$^1$Graduate School of Engineering and Resource Science, Akita University,\\
    Tegata, Akita City 010-8502, Japan\\
$^2$Research Center for Nuclear Physics, Osaka University,\\ 
	10-1 Mihogaoka, Ibaraki, Osaka 567-0047, Japan
}

\abst{%         %This abstract is neglected when [addenda] or [errata].
Resonances observed in the $^{28}\rm Si+{}^{28}Si$ collision are studied 
by the molecular model. 
In the preceding paper, it is clarified that 
at high spins in $^{28}\rm Si+{}^{28}Si$ (oblate-oblate system), 
the stable dinuclear configuration of the system is equator-equator 
touching one, 
and that the axially asymmetric shape of the stable 
configuration of $^{28}\rm Si+{}^{28}Si$ gives rise to 
a wobbling motion ($K$-mixing). 
There, the normal modes around the equilibrium have been solved 
and various excited states have been obtained. 
Those states are expected to be the origin of 
a large number of resonances observed. 
Hence their physical quantities are analyzed theoretically. 
The results are compared with the recent experiment performed 
in Strasbourg and turn out to be in good agreement with the data.
Disalignments between the orbital angular momentum and 
the spins of the constituent $^{28}\rm Si$ nuclei 
in the resonance state are clarified. 
Moreover the analyses of the angular correlations indicate 
characteristic features for each normal-mode excitation. 
Thus it is possible to identify the modes, and a systematic 
experimental study of angular correlation measurements is desired. 
}

\PTPindex{223}  %Input the subject index(es) of your paper, 
\begin{document}
\maketitle

%%%%%%%%%%%%%%%%%%%%%%
\section{Introduction}
%%%%%%%%

Intermediate resonances observed in heavy-ion scattering have offered 
intriguing subjects in nuclear physics.\cite{BettsC-a, BettsC-b} %%[1,2]  
In the preceding paper,\cite{UeNewI} %%[3]
(hereafter referred as paper~I), the authors have studied dinuclear 
molecular structure of the $^{28}\rm Si+{}^{28}Si$ system. 
In the present paper, we investigate physical quantities by using the 
molecular wave functions and compare the results with the experiments. 

Betts et al. firstly observed a series of resonance-like enhancements 
at $\theta_{\rm cm}=90^\circ$ in elastic scattering of 
$^{28}\rm Si+{}^{28}Si$, in the energy range from $E_{\rm lab}=101$MeV 
to $128$MeV with broad bumps of about $2$MeV width.
They gave spin assignments of $J=34 - 42$ by the Legendre-fits 
to the elastic angular distributions for each bump, which correspond 
to the grazing partial waves.\cite{BettsPRL1,BettsPL} 
They further closely investigated angle-averaged excitation functions 
for the elastic and inelastic scatterings, 
and found, in each bump, several sharp peaks correlating among 
the elastic and inelastic channels.\cite{BettsPRL2,SainiBetts} 
The total widths of those resonances are about $150{\rm keV}$, and the 
inelastic decay strengths are enhanced and stronger than the elastic one,
which suggests that they are special eigenstates of the compound system. 
Similar sharp resonance peaks are observed by Zurm\"uhle et al. in the 
$^{24}\rm Mg+{}^{24}Mg$ system.\cite{Zurm}
In those systems, the decay widths of the elastic and inelastic channels 
up to high spin members of the $^{24}\rm Mg$ or $^{28}\rm Si$ ground 
rotational band exhaust about $30\%$ of the total widths, whereas those 
into $\alpha$-transfer channels are much smaller.\cite{Saini,Beck2000}
These enhancements of symmetric-mass decays strongly suggest 
dinuclear molecular configurations for the resonance states. 
It is also noted that the widths of the elastic channel are rather small, 
for example, a few keV, 
which is quite different from high spin resonances in lighter systems 
such as 
${}^{12}\rm {C} +{}^{12}\rm {C}$ and 
${}^{16}\rm {O} +{}^{16}\rm {O}$.

Taking into account the difference from resonances in lighter systems 
and the level density of the sharp resonances observed, 
we have developed a new dinucleus-molecular model 
for the high spin resonances in the $^{24}\rm Mg +{}^{24}Mg$ 
and $^{28}\rm Si +{}^{28}Si$ systems,\cite{Ue89,Ue93,Ue94,UeSuppl} 
in which two incident ions are supposed to form a united composite system. 
It rotates as a whole in space with internal degrees of freedom 
originating from interaction of the deformed constituent ions.  
This is in contrast with the viewpoint of 
the band crossing model.\cite{BCMSuppl}

%%%%%%%%%%%%%%%%%%%%%%%%%%%%%%%%%%%%%%%
%%%%%%%%%%%%%%%%%%%%%%%%%%%%%%%%%%%%%%%
\begin{wrapfigure}{r}{6.6cm}
%%  \figurebox{60mm}{9cm}
\centerline{\includegraphics[%width=WIDTH cm,
                             height=8.6 cm]
                             {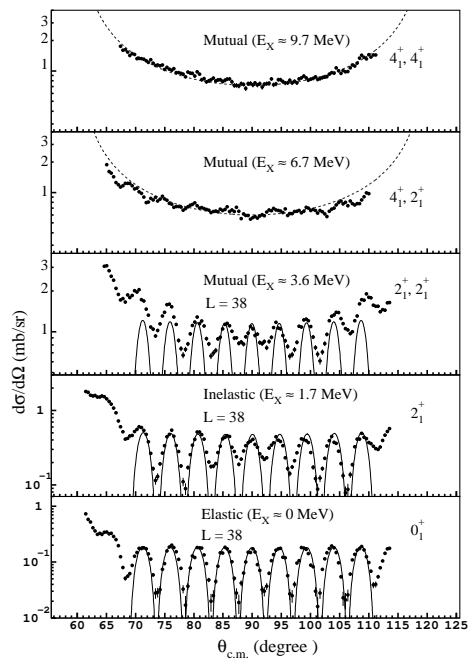}}
\caption{Experimental angular distributions for the elastic and 
inelastic scattering for $^{28}\rm Si+{}^{28}Si$ 
at $E_{\rm CM}=55.8$MeV.\cite{Beck2000,Nouicer99} 
Solid curves show $L=38$ Legendre fits for comparison.}
\label{fig:1}
\end{wrapfigure}
%%%%%%%%%%%%%%%%%%%%%%%%%%%%%%%%%%%%%%%
%%%%%%%%%%%%%%%%%%%%%%%%%%%%%%%%%%%%%%%
%

In the paper~I,\cite{UeNewI} we have applied the model also 
to $^{28}\rm Si + {}^{28}Si$, 
though there is a distinct difference between prolate-prolate  
and oblate-oblate systems. 
In the former, the stable configuration is pole-to-pole one 
due to the prolate deformation of $^{24}\rm Mg$, while in the latter, 
it is the equator-to-equator configuration due to the oblate deformation 
of $^{28}\rm Si$.  Therefore, the former composite system is axially 
symmetric in the equilibrium, while the latter is triaxial.  
Then, in the latter, strong $K$-mixing is kinematically induced, 
and results in a wobbling motion.\cite{Wobb}%%[21]%
Thus we extend the molecular model to the axially-asymmetric 
configurations. 
In practice, we do not treat Coriolis terms in the hamiltonian 
explicitly, but we diagonalize the hamiltonian of the asymmetric 
rotator due to the triaxiality to obtain new low-lying states.  
They are expected to correspond to the sharp 
resonance peaks within each bump of the grazing $J$.  
Therefore, it appears that the present model is promising for the sharp 
high spin resonances observed in $^{28}\rm Si +{}^{28}Si$, 
but it is difficult to proceed further, i.e., 
to assign which mode of excitation corresponds to which peak observed, 
due to less detailed experimental information.  Fortunately, 
there are exceptional data on the resonance at $E_{\rm cm}=55.8$MeV 
in $^{28}\rm Si +{}^{28}Si$, 
as we will discuss in detail in the present paper.

A new facet of the resonance states has been explored by  
the $^{28}\rm Si+{}^{28}Si$ scattering experiment on the resonance 
at $E_{\rm cm}=55.8$MeV at IReS Strasbourg.\cite{Nouicer99}
Figure~1 shows those elastic and inelastic angular distributions, 
where the solid lines in the lower three panels show $L=38$ Legendre fits.
The oscillating patterns in the elastic and inelastic channels 
$2^+$, $(2^+,2^+)$ are found to be in good agreement with $L=38$, 
which suggests $L=J=38$ dominance in the resonance, 
namely, {\em disalignments} of the fragment spins and 
the orbital angular momentum. 
$\gamma$-rays emitted from the emerging fragments, the excited 
$^{28}\rm Si$ nuclei, have been also measured with $4\pi$ detectors 
in coincidence with two  $^{28}\rm Si$ fragments 
detected at $\theta_{\rm cm} = 90^\circ$.
Those angular correlation data show characteristic $"m=0"$ patterns 
in the axis normal to the reaction plane, which suggests that the 
fragment spins are in the scattering plane and is consistent with 
the disalignments observed in the fragment angular distributions.
The disalignments have never been observed before in heavy-ion reactions, 
which indicates that the $^{28}\rm Si+{}^{28}Si$ system is completely 
different from $^{12}\rm C+{}^{12}C$ etc. and even from 
the $^{24}\rm Mg+{}^{24}Mg$ system which exhibit spin 
alignments.\cite{Trombik84,Konnerth85,Mattis87,KonnCluster88,Wuosmaa87} 
It, thus, is worth to emphasize here that the angular correlation data 
provide crucial information on configurations of the constituent nuclei 
in the resonance state, which could not be obtained otherwise. 
Therefore, we concentrate our analyses on the data on the resonance 
$E_{\rm cm}=55.8$MeV in the following. 
With the new data on the angular correlations, we are now able to select 
"which mode is really a candidate for the resonance".

For this purpose, we have developed a method of analysis 
for "particle-particle-$\gamma$ angular correlations", 
in the framework of $R$-matrix theory.\cite{KapurPeierls,LainThomas}
The results obtained by the analysis of the angular correlations 
show that $K$-mixing due to 
the rotational motions of the triaxially-deformed system 
is found to be indispensably important for understanding 
the $"m=0"$ dominance of the angular correlations, 
which brings novel aspects of "wobbling motion"\cite{Wobb}%%[21] 
in the resonances, or in the reaction mechanism.

In \S2,    %%%%section 2, 
formulation of the $R$-matrix theory is briefly reminded, 
and expressions of angular correlations are given in the framework. 
In \S3,    %%%%section 3, 
theoretical analyses and comparisons with the experimental data 
are made. 
In \S4,    %%%section 4, 
we discuss the mechanism of the spin disalignments 
and the observation conditions of the resonances.
Conclusions are given in \S5.   %%%%section 5.

\section{Formulation on decay properties and angular correlations}
%%%%%%%%%%%%%%%%%%%%%% Start your paper from here.

For calculations of the decay properties of the molecular 
resonance states, we prepare necessary expressions, 
based on $R$-matrix theory,\cite{LainThomas} 
since we have the energy spectra and the wave functions 
of the molecular states obtained in the paper~I.\cite{UeNewI}

\subsection{$R$-matrix formalism}

We write the channel spin wave functions $\psi$ as 
\begin{equation}
\psi_{\tau (I_1 I_2) I M_I}=  \sum_{M_1 M_2} (I_1 I_2 M_1 M_2| I M_I)
   \,  \chi_{\tau_1 I_1 M_1}  \cdot  \chi_{\tau_2 I_2 M_2},
\label{eq:2.1-1}
\end{equation}
where $(I, M_I)$ denote the channel spin, and $\chi_{\tau_i I_i M_i}$ 
describe the states of the separated constituent nuclei. 
With $(\tau_i I_i M_i)$,  $\tau_i$ specifies the internal state 
of the {\it i}-th constituent nucleus, and  $(I_i M_i)$ denote 
the spin quantum numbers. 
Since we consider excitations to the members of the ground-state 
band of $^{28}\rm Si$ nucleus, $\tau_i$ is taken to be the same as that 
of the ground state, and thus we omit it in the following descriptions. 
Complete channel wave functions are given as  
\begin{equation}
\Psi_{c,I M_I lm} =  {u_{cl}(r_c) \over \sqrt{v_c} r_c} 
                            Y_{lm}(\theta, \phi) 
                            \psi_{(I_1 I_2) I M_I},
\label{eq:2.1-2}
\end{equation}
where $u_{cl}(r_c)$ denotes the radial wave function between 
two constituent nuclei in the channel $c$, 
which specifies  $(I_1,I_2)$, 
$l$ being the orbital angular momentum. 
For the analyses of resonance states with the total spin $(J, M)$, 
hereafter we use generalized channel wave functions in the 
representation $\{(I_1 I_2) I l , J M \}$  as 
\begin{equation}
 \hbox{\calg Y}_{c I l JM}
    = \sum_{M_I, m} (I l M_I m | J M)  \psi_{(I_1 I_2) I M_I}
      Y_{lm}(\theta, \phi) .
\label{eq:2.1-3}
\end{equation}
For the radial wave function $u_{cl}(r_c)$, the asymptotic functions 
are given with 
\begin{eqnarray}
u^{(-)}_{cl}  &= e^{i\sigma_{cl}}(G_{cl} - iF_{cl}), 
\nonumber \\  
u^{(+)}_{cl}  &= e^{-i\sigma_{cl}}(G_{cl} + iF_{cl}), 
\label{eq:2.1-4}
\end{eqnarray}
where $u^{(-)}_{cl}$ and $u^{(+)}_{cl}$ denote incoming and 
outgoing waves, respectively, 
with $\sigma_{cl}$ being Coulomb phase shift. 
$F_{cl}$ and $G_{cl}$ denote the regular and irregular 
Coulomb wave functions.

Specifying the incident channel with two $^{28}\rm Si$ nuclei 
in the ground state, with $l=J$ and $m=0$ with $z$-axis 
parallel to the beam direction, the wave function 
of the system for the external region is written as
\begin{equation}
\Psi   \sim {u^{(-)}_{cl} \over \sqrt {v_c} r_c} 
           \hbox{\calg Y}_{c I l JM} 
      -\sum_{c'I'l'} U_{c'I'l',cIl} 
          {u^{(+)}_{c'l'} \over \sqrt {v_{c'}} r_{c'}}
                       \hbox{\calg Y}_{c'I'l'JM} ,      
\label{eq:2.2}
\end{equation}
where $c$ denotes the initial channel $(I_1=I_2=0)$, 
and $c'$ denotes the final channel, respectively. 
By using $R$-matrix formula,\cite{LainThomas} %%[19,20], 
we obtain the collision matrix, 
\begin{equation}
U_{c'I'l',cIl} = 
  {u^{(-)}_{cl}(k_c,a_c) \over u^{(+)}_{cl}(k_c,a_c)} \delta_{c'c}
  -i\sum_{\lambda} {u^J_{\lambda c'I'l'}{\tilde u}^{J*}_{\lambda cIl}
      \over N^J_\lambda (E-W^J_\lambda)},
\label{eq:2.3}
\end{equation}
where $a_c$ denotes channel radius.
The second term of Eq.~(\ref{eq:2.3}) is a sum of the contributions 
from the resonance states $\lambda$.   $W^J_\lambda$ is a resonance pole, 
i.e., 
$W^J_\lambda= E^J_\lambda -{i \over 2}\Gamma^J_\lambda$, 
and $N^J_\lambda$ corresponds to a factor for the normalization 
of the resonance state $\lambda$, which is close to unit.
$u^J_{\lambda cIl}$ are defined by 
\begin{equation}
u^J_{\lambda cIl}= 
{\sqrt{2 k_c a_c} \over u^{(+)}_{cl}(k_c,a_c)} 
        \gamma^J_{\lambda cIl},
\label{eq:2.4}
\end{equation}
with the reduced widths $\gamma^J_{\lambda cIl}$ from the amplitudes of 
the resonance states in the channel $c$, and $\gamma^J_{\lambda cIl}$ 
are given by 
\begin{equation}
\gamma^J_{\lambda cIl} = 
        \Biggl( {\hbar^2 \over 2\mu_c a_c}\Biggr)^{1/2}
   {1 \over a_c} \int 
       \hbox{\calg Y}^*_{c Il JM}  \cdot  \Psi^{JM}_\lambda dS,
\label{eq:2.5}
\end{equation}
where $\Psi^{JM}_\lambda$ denotes decaying resonance states, 
$S$ being the surface at the channel radius $a_c$.
Note that notations ${\tilde u}^{J}_{\lambda cIl}$ 
correspond to the definition of Eq.~(\ref{eq:2.4}), 
but with the incoming waves $u^{(-)}_{cl}$ instead of $u^{(+)}_{cl}$, 
as is usual in the $R$-matrix theory.
Calculations of Eq.~(\ref{eq:2.5}) need relations between 
molecular wave functions and the generalized channel wave functions, 
which are described in Appendix A.

For inelastic scattering, $U_{c'I'l',cIl}$ is given 
only by the second term.
We apply one level approximation, i.e., we consider the energy 
region close to a resonance level $\lambda$ and replace the sum of 
the second term of Eq.~(\ref{eq:2.3}) 
by one term with the resonance state $\lambda$, the effect of the 
neglected terms being taking into account by the modification of 
the hard sphere scattering phase shift as is explained later, 
and then we have
\begin{equation}
U_{c'I'l',cIl}= -i {u^J_{\lambda c'I'l'} {\tilde u}^{J*}_{\lambda cIl} 
                \over      N^J_\lambda (E-W^J_\lambda)}.
\label{eq:2.6}
\end{equation}
We obtain cross sections for the reaction $c' \ne c$,
\begin{equation}
\sigma_{c'c} = g_J {\pi \over k^2_c} 
               \sum_{II' ll'} |U_{c'I'l',cIl}|^2
             = g_J {\pi \over k^2_c} 
               \sum_{II' ll'} 
        { |u^J_{\lambda c'I'l'}|^2 |{\tilde u}^{J*}_{\lambda cIl}|^2 
   \over 
  |N^J_\lambda|^2 \{(E-E^J_\lambda)^2 +{1 \over 4}\Gamma^2_\lambda \} },
\label{eq:2.7}
\end{equation}
where $ g_J= {(2J+1)/(2I_1 +1)(2I_2 +1) }$ is the statistical factor.
Now we define partial widths in the channel $c'$ 
of the level $\lambda$ by
\begin{equation}
\Gamma^J_{\lambda c'}= 
           \sum_{I'l'} |u^J_{\lambda c'I'l'}|^2 = 
           \sum_{I'l'} 2P_{c'l'} |\gamma^J_{\lambda c'I'l'}|^2 ,
\label{eq:2.8}
\end{equation}
where $P_{cl}$ denote so-called penetration factor given by
\begin{equation}
P_{cl} = {k_c a_c \over G_{cl}^2 + F_{cl}^2},
\label{eq:2.9}
\end{equation}
and the total width of the level $\lambda$ is obtained 
by the sum of the partial widths,
\begin{equation}
 \Gamma^J_{\lambda}= \sum_{c} \Gamma^J_{\lambda c}.
\label{eq:2.10}
\end{equation}
Hence, with $|N^J_\lambda| \sim 1$, 
we obtain the Breit-Wigner one-level formula,
\begin{equation}
\sigma_{c'c} \sim  g_J  
              {\pi \over k^2_c} 
    {    \Gamma_{\lambda c'} \Gamma_{\lambda c} 
     \over 
         \{(E-E^J_\lambda)^2 + {1 \over 4}\Gamma^2_\lambda \}  }.
\label{eq:2.11}
\end{equation}
In particular, for "on resonance", i.e., for $E=E^J_\lambda$, 
we obtain
\begin{equation}
U_{c'I'l', cIl}(E_\lambda) = i^{l-l'}e^{i(\sigma_{cl}+\sigma_{c'l'})}
     {e^{-i\phi'_{c'l'}} (-2\sqrt{2P_{c' l'}} \gamma_{c'I'l'}
           \sqrt{2P_{c l}} {\tilde \gamma}^*_{cIl}) e^{-i\phi_{cl}} 
        \over N^J_\lambda \Gamma^J_\lambda  },
\label{eq:2.12}
\end{equation}
where the indications of $J$ and $\lambda$ 
for the resonance state are omitted.
$\phi_{cl}$ denote hard-sphere scattering phase shifts, which are 
defined by $\tan \phi_{cl} ={ F_{cl}(k_c, a_c) / G_{cl}(k_c, a_c) }$. 
Note that in the expression Eq.~(\ref{eq:2.10}) for the total width, 
the channels $c$ should be taken generally not only for channels 
such as the elastic, the single and mutual $2^+$ but also for those 
associated with higher fragment spins. 
Furthermore experimentally it is suggested that the contributions 
from symmetrical-dinucleus channels exhaust only about one third 
of the total widths.\cite{Saini} %%[9].
Therefore we practically adopt the total width 
$\Gamma^J_\lambda = 150{\rm keV}$ for Eq.~(\ref{eq:2.12}) which is 
suggested by the experiments.\cite{BettsPRL2} %%[6)]. 

To calculate $\gamma^J_{\lambda cIl}$ by Eq.~(\ref{eq:2.5}), 
we adopt the model wave functions 
obtained by the bound-state approximation for $\Psi^{JM}_\lambda$.
Relations between the molecular wave functions and the channel wave 
functions necessary for the calculations are given in Appendix A.
In order to avoid large dependence on the channel radius 
due to gaussian damping of the model radial wave function, 
which is due to the harmonic approximation, 
we have corrected tails of the radial wave functions 
by smoothly connecting $G_{cl}$'s in the concerning channels. 
At $E_{\rm cm}=55.8$MeV the distances of the connecting points are 
in the range of $R_c = 8.3 - 8.8$fm, and each channel radius $a_c$ 
is taken to be larger than the connecting distance, 
which means pure Coulomb interaction is assumed outside $a_c$.  
From $R_c$ to $a_c$, internal waves in the channel $(c, l)$ of 
$ \Psi^{JM}_\lambda$ are assumed to behave to be proportional to 
$( G_{cl} + i F_{cl})$, like the outgoing waves.
Hence the channel radius dependence is removed except for the 
small renormalization effect due to $N^J_\lambda$, which is defined by 
$N^J_\lambda = \int {\tilde \Psi}^{JM*}_\lambda \Psi^{JM}_\lambda d\tau $ 
with the integration inside $a_c$. 
By using wave functions normalized in the bound state approximation,  
$N^J_\lambda$ is usually real and close to unit. 
In detail, a small difference occurs from modified radial tails 
around $a_c$, which, of course, depends on $E_{\rm cm}$ and $a_c$.
The difference in each channel appears to be a few percent 
at the maximum, and 
for simplicity we have approximated the value of $N_\lambda$ 
to be equal to one in the present calculations.

For the elastic scattering in which various partial waves 
contribute from $l=0$ to larger than grazing partial ones, 
it is better to rewrite Eqs.~(\ref{eq:2.2}) and (\ref{eq:2.3}) 
explicitly with $I_1=I_2=0$ and  $l$'s.
We have 
\begin{equation}
\psi = 
          {1 \over \sqrt {v} k r} 
            \sum_l (2l+1) i^l 
            \bigg[ e^{i\sigma_{l}} F_{l}
      + {i \over 2}(e^{2i\sigma_{l}} -U_{l}) u^{(+)}_{l} \bigg] 
        P_l(\cos\theta),
\label{eq:2.13}
\end{equation}
with $c$ being dropped in the notations, 
and the differential cross section is given by 
\begin{equation}
{d \sigma \over d\Omega} = 
          \bigg| f_C(\theta) 
          + {i \over 2k }  \sum_l (2l+1)  
          (e^{2i\sigma_{l}} -U_{l}) P_l(\cos\theta) \bigg|^2,
\label{eq:2.14}
\end{equation} 
where $f_C(\theta)$ denotes the Coulomb scattering amplitude, 
and $U_{l}$ denotes the collision matrix for the $l$-partial waves.
As for the collision matrix, we modify Eq.~(\ref{eq:2.3}) as follows. 
Firstly in order to do one-level approximation, 
we divide the second term of Eq.~(\ref{eq:2.3}), a sum of the 
contributions from the resonance states $\lambda$, into two groups, 
\begin{equation}
U_{l} = {u^{(-)}_{l}(k,a_c) \over u^{(+)}_{l}(k,a_c)} 
     -i\sum_{{\lambda'} \ne \lambda} { u^J_{\lambda'} 
                                   {\tilde u}^{J*}_{\lambda'}
      \over N^J_{\lambda'} (E-W^J_{\lambda'})}
    -i{ u^J_{\lambda}  {\tilde u}^{J*}_{\lambda}
                  \over N^J_\lambda (E-W^J_\lambda)},
\label{eq:2.15}
\end{equation}
where the third term is the contribution from the resonance under 
consideration, and the second term is those from the other levels. 
In general, the second term includes not only the contributions from 
the resonance states with the $^{28}\rm Si+{}^{28}\rm Si$ configurations,  
but also those from the resonance states with the different configurations 
and from the compound nucleus states. 
Since the contributions to the total widths from the symmetrical 
dinuclear channels exhaust only about one third, as already mentioned, 
the contributions from the other channels are significant. 
The effects of the second term brought from the other decay channels 
and the fusion channels in average are expected to appear as absorption 
in the elastic channel. 
Thus we replace the sum of the first and second terms of the r.h.s. of 
Eq.~(\ref{eq:2.15}) 
with nuclear scattering amplitudes which is phenomenologically proposed 
by the study of heavy-ion scattering,\cite{Frahn} %%[31], 
and obtain the collision matrix for the $l$-partial waves as 
\begin{equation}
U_{l} = A_l    -i{ u^J_{\lambda }  {\tilde u}^{J*}_{\lambda }
                       \over N^J_\lambda (E-W^J_\lambda)} ,
\label{eq:2.16}
\end{equation}
where $A_l$ denotes the elastic scattering amplitude 
for the $l$-partial wave.
Supposing that the lower partial waves than the grazing one with 
$l_{gr} \sim J$ would be strongly absorbed as usual in the heavy-ion 
scattering, we assume reflection coefficients to be gradually 
increasing as $l$ comes to be over the grazing one, i.e., 
we apply smooth cutoff model,\cite{McIntyre}   %%[26],  
\begin{equation}
A_l  =  { e^{2i(\sigma_{l}+\delta_{l})} 
                 \over 1+ \exp[-(l-(J+\Delta))/\Delta_l] } , 
\label{eq:2.17}
\end{equation}
with nuclear phase-shifts being
$\delta_l = 2\delta /\{1+ \exp((l-(J+\Delta))/\Delta_l)\}$.
We put $\Delta=2$, $\Delta_l=3.3$ and $\delta=0.1$ 
to reproduce the experimental data 
in the region of $\theta=20^\circ - 60^\circ$ where the 
differential cross section is very rapidly decreasing 
as $\theta$ increases.\cite{BettsC-a, BettsC-b} %%[1,2].

Finally it is noted that for the collisions of identical particles, 
scattering cross sections are modified due to the symmetry between 
the particles. For the final channels with identical particles 
(the cases of $I_1 =I_2$), the expression of 
the differential cross sections is given by 
$|(1/\sqrt2) \{ f_{c'}(\theta, \phi)
       +f_{c'}(\pi-\theta, \phi+\pi) \}|^2 \times 2$, 
where $f_{c'}$ denote the amplitudes of the outgoing waves on the 
spherical surface. The factor $(1/\sqrt2)$ is due to the normalization 
of the incident channel and the multiplied factor 2 corresponds to the 
detections of the scattered particle and the recoil particle. 
Thus the scattered partial waves are allowed only for $l=even$, and 
the cross sections $\sigma_{c'c}$ in Eq.~(\ref{eq:2.11}) are 
modified as well, by factor 2. 
Furthermore careful calculations of $d\sigma(\theta) / d\Omega$ 
by using Eqs.~(\ref{eq:2.14}) and (\ref{eq:2.16}) are necessary. 
The resonance term is obtained with the wave functions 
with boson symmetry. 
On the other hand the other amplitudes $f_C(\theta)$ etc. are not 
symmetrized. Thus the differential cross section for the elastic 
scattering of the spinless identical particles is given as 
\begin{equation}
{d \sigma \over d\Omega} = 
          \bigg| {1 \over \sqrt2}\{f(\theta)+f(\pi-\theta)\} 
       -{1\over 2k}(2J+1) { u^J_{\lambda }{\tilde u}^{J*}_{\lambda }
                \over N^J_\lambda (E-W^J_\lambda)} P_J(\cos\theta)
          \bigg|^2 \times 2,
\label{eq:2.14-2}
\end{equation}     
with the amplitude $f(\theta)$ by the Coulomb and nuclear scattering 
except for the single resonance term, 
\begin{equation}
 f(\theta)=  f_C(\theta) 
            +{i \over 2k }  \sum_l (2l+1)  
           (e^{2i\sigma_l} -A_l) P_l(\cos\theta) .
\label{eq:2.14-3}
\end{equation}     

\subsection{Fragment-fragment-$\gamma$ angular correlation}

For the calculations of the angular correlations, we need the 
scattering amplitudes in magnetic substates $(m_1 , m_2)$ 
representation, which are connected to the subsequent 
$\gamma$-ray emission process. The amplitudes for the mutual 
$2^+$ excitation are given by
\begin{eqnarray}
X_{m_1 m_2}(\boldsymbol{k}',\boldsymbol{k}) =  {2\pi \over ik} 
                  \sum_{I'l'm'} & (22m_1m_2|I'm_1+m_2)(I'l'm_1+m_2 \ m'|JM)
\nonumber \\  
            & \times  
	             U^J_{I'l'} Y^*_{JM}({\hat k})  Y_{l'm'}({\hat k}')  ,
\label{eq:2.18}
\end{eqnarray}
where $\boldsymbol{k}$ and $\boldsymbol{k}'$ denote the initial and final 
relative momenta between two $^{28}\rm Si$ nuclei, respectively, 
and the incident (elastic) channel assignment for the collision matrix 
is omitted. 
Here we have assumed a single resonance with a total angular momentum $J$. 
For the single excitation, of course, we have a similar expression 
as the above, 
by putting $I_2 =0$ and $m_2=0$ into the CG coefficients 
of the mutual-channel spin coupling in Eq.~(\ref{eq:2.18}). 
The transition amplitudes for the $\gamma$-ray emissions from the polarized 
nuclei are discussed by several authors.\cite{Rybicki} %%[33]. 
For two photon emissions from the mutual excitation, 
the amplitudes are proportional to the scattering amplitudes and 
the strength of the photon emission as 
\begin{equation}
A^{\sigma_1 \sigma_2}_{I_1 I_2}  \sim
\sum_{m_1 m_2} X_{m_1 m_2}(\boldsymbol{k}',\boldsymbol{k})
                (00|H_{\sigma_1} | I_1 m_1)
                (00|H_{\sigma_2} | I_2 m_2), 
\label{eq:2.19}
\end{equation}
where $(00|H_{\sigma} | I_i m_i)$ denote the transition matrices 
for $\gamma$-ray emissions, which give transition rates, i.e., 
by well-known perturbation theory, 
$2\pi/\hbar \cdot |(00|H_{\sigma} | I_1 m_1)|^2 \rho d\Omega$ 
for the direction $d\Omega$ of the emission, with $\rho$ being the level 
density of the final states. 
In the case of the present angular correlations, photons are rapidly 
emitted with the half life of $700$fs, which means almost all the first 
$2^+$ states of $^{28}$Si nuclei finish their transitions into the ground 
states just after the decays of the resonance compounds. 
Thus the problem which we will discuss is not 
the magnitudes of the transition rates but angular distributions 
of the $\gamma$-ray intensities over $4\pi$-detectors, 
which were measured by Eurogam Phase \RMNb .
The $\gamma$-ray intensity distributions are given by
\begin{eqnarray}
 (00|H_{\sigma}(I m) | I m)  & \sim 
                   D^I_{m \sigma}(\phi_\gamma, \theta_\gamma,0)
                              (I I m -m |00)(0||T(I)||I)
\nonumber \\ 
          & \sim  {(-1)^{I-m}  \over \sqrt{2I+1} } 
                 d^I_{m \sigma}(\theta_\gamma)e^{-im\phi_\gamma},
\label{eq:2.20}
\end{eqnarray}
where $\sigma$  denote 
right/left-hand circular polarizations of the emitted $\gamma$-rays, 
i.e., $\sigma = \pm 1$.  
After sum over them for the square of the absolute values 
of Eq.~(\ref{eq:2.19}), we obtain $\gamma$-ray angular correlations,
\begin{eqnarray}
W_{I_1 I_2}( \theta_1,\phi_1; \theta_2,\phi_2; \boldsymbol{k}',\boldsymbol{k})
   \equiv  &  \sum_{\sigma_1 \sigma_2} |A^{\sigma_1 \sigma_2}_{I_1 I_2}|^2
\qquad\qquad\qquad\qquad\qquad%%\qquad\qquad
\nonumber \\  %%\cr
    =  & \sum_{\sigma_1 \sigma_2 m_1 m_2 m'_1 m'_2} 
               X_{m_1 m_2}(\boldsymbol{k}',\boldsymbol{k}) 
               X^*_{m'_1 m'_2}(\boldsymbol{k}',\boldsymbol{k})
\nonumber \\  %%\cr
  & \qquad \qquad \times  (00|H_{\sigma_1} | I_1 m_1) 
                                 (00|H_{\sigma_1} | I_1 m'_1)^*
\nonumber \\  %%\cr
  & \qquad \qquad \times  (00|H_{\sigma_2} | I_2 m_2) 
                                 (00|H_{\sigma_2} | I_2 m'_2)^*,
\label{eq:2.21}
\end{eqnarray}
where $(\theta_i, \phi_i)$ denote the directions of the emitted photons.
In the experiment, only one of the two emitted photons is detected 
in most cases, even with the Eurogam.\cite{Beck-priv} %%[28]. 
Therefore we take up the distribution of one photon 
after averaging over the distribution of the other photon. 
By taking average over the angles of the second photon 
$(\theta_2, \phi_2)$, the last line of Eq.~(\ref{eq:2.21}) is reduced to 
$\delta_{m_2 m'_2}$ due to the integration for the phase factor 
$e^{-i(m_2-m'_2)\phi_2}$ appearing from Eq.~(\ref{eq:2.20}).
Hence the $4\pi$ $\gamma$-ray intensity distribution is given by 
\begin{eqnarray}
W_{I_1 I_2}( \theta_1,\phi_1;\boldsymbol{k}',\boldsymbol{k}) 
                  \sim  \sum_{m_1 m'_1 } 
      & \Big\{ \sum_{m_2 } 
       X_{m_1 m_2}(\boldsymbol{k}',\boldsymbol{k}) 
       X^*_{m'_1 m_2}(\boldsymbol{k}',\boldsymbol{k}) \Big\}
\nonumber \\  
& \times (00|H_{\sigma_1}| I_1 m_1)(00|H_{\sigma_1}| I_1 m'_1)^* ,
\label{eq:2.22}
\end{eqnarray}
where sum over $m_2$ is associated with the average on one photon.

%%% Ang. Correlation pattern(2 figures) %%%%%%%%%%%%%%%%%%%%%%%%%%%%
\begin{figure}[b]  %FIG-2, -3(W_m, maximum-aligned config.) one set
  \parbox[t]{\halftext}{%   
  \centerline{\includegraphics[%%width=WIDTH cm,
                             height=7.2 cm]   %%% 6.3 cm] %%  7.5 cm]
                             {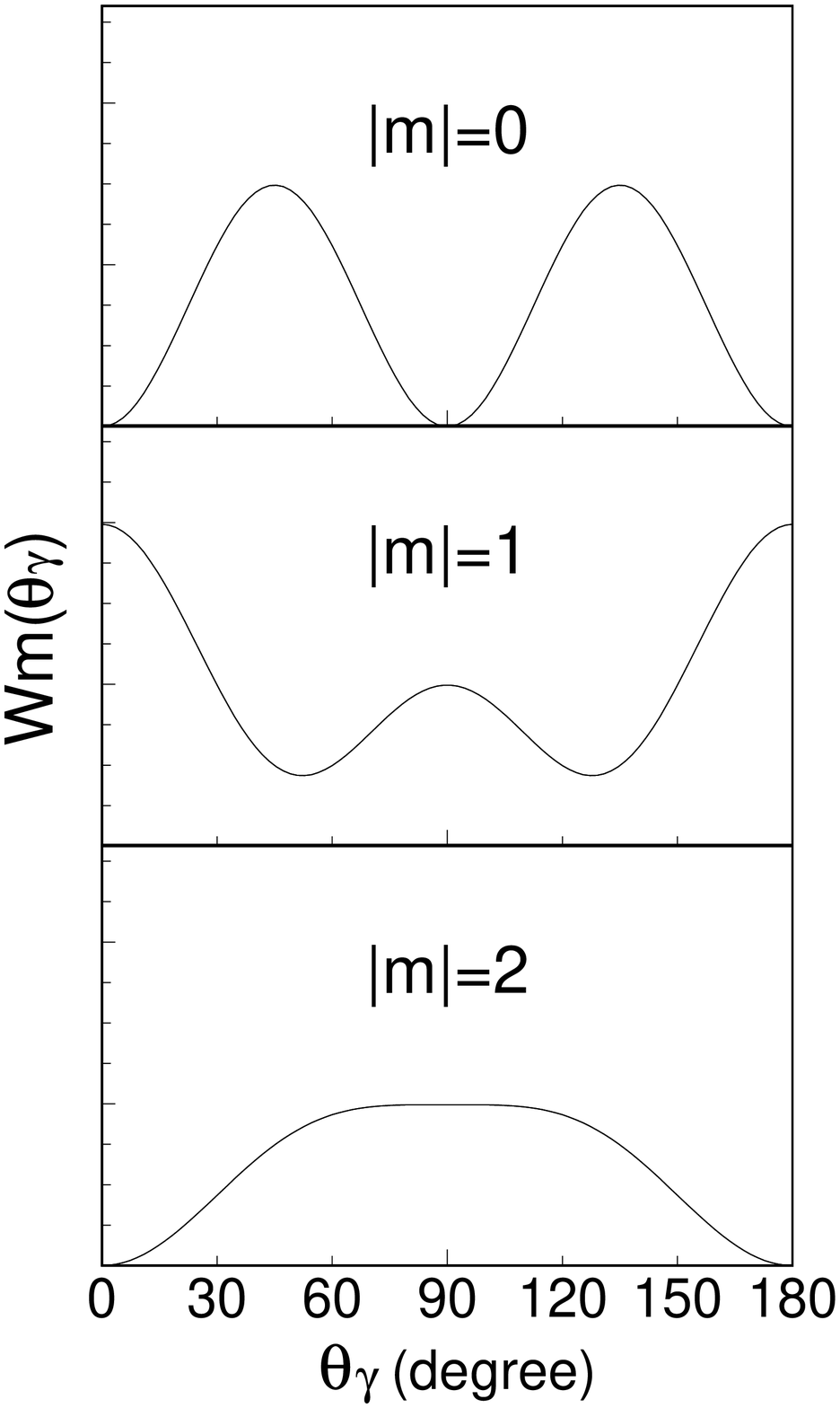}}
\caption{$\gamma$-ray intensity distributions $W_m(\theta_\gamma)$ 
of the pure magnetic substates for $m=0, \pm 1$ and $\pm 2$. 
From the top panel, $m=0$, $|m|= 1$ 
and $|m|= 2$ are shown, respectively. }
}
  \label{fig:2}
  \hfill
  \parbox[t]{\halftext}{%   %\def\halftext{.471\textwidth}
  %%\figurebox{60mm}{70cm}
  \centerline{\includegraphics[%%width=WIDTH cm,
                             height=7.2 cm]   %%% 6.3 cm] %%  7.5 cm]
                             {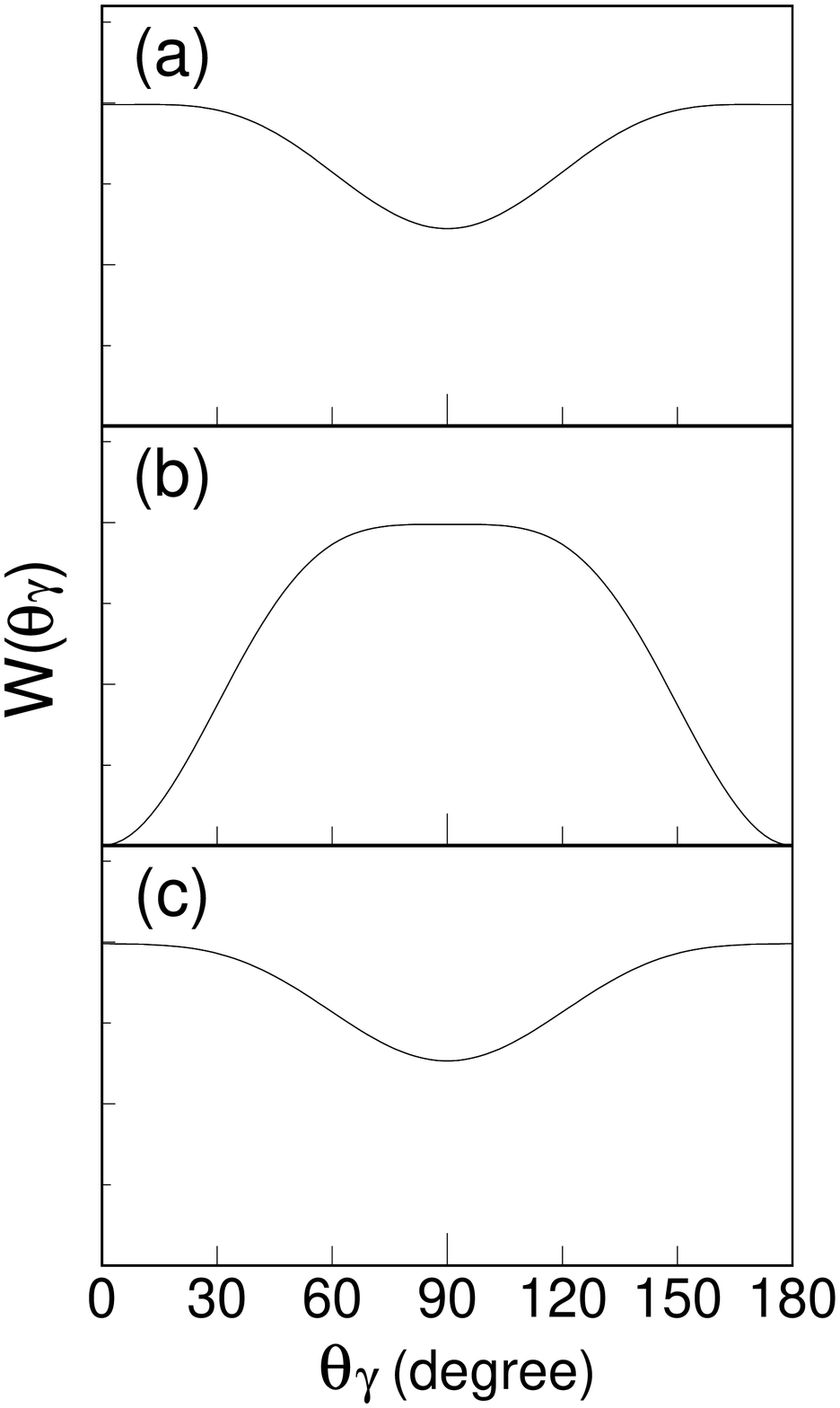}}
\caption{$\gamma$-ray intensities for 
particle-particle-$\gamma$ angular correlations 
with the maximum-aligned configuration $J=38$ with $l'=34(I'=4)$ 
of the mutual $2^+$ channel. 
For the quantization axes in three panels, see text. } 
}
  \label{fig:3}
\end{figure}    %%% FIG-2, FIG-3 END

Now with the expression of Eq.~(\ref{eq:2.22}), we are able to compare 
the theoretical results with the experimental $4\pi$ $\gamma$-ray 
distribution. As is usually done, it is useful to determine 
the contributions from the magnetic substates.\cite{Wuosmaa87} 
In order to obtain the probabilities in the magnetic substates 
of the fragments, we integrate Eq.~(\ref{eq:2.22}) over $\phi_1$ 
(averaged around the $z$-axis). 
Then only diagonal elements with $m_1=m'_1$ appear 
because of the phase $e^{-i(m_1 -m'_1)\phi_1}$ from Eq.~(\ref{eq:2.20}). 
Thus the normalized angular correlations are expressed as follows: 
\begin{equation}
W(\theta_\gamma) = \sum_m  P_m W_m(\theta_\gamma),
\label{eq:2.23}
\end{equation} 
with
\begin{equation}
P_m \sim   \sum_{m_2} |X_{m m_2}(\boldsymbol{k}',\boldsymbol{k})|^2,
\label{eq:2.24}
\end{equation}
and 
\begin{equation}
W_m(\theta_\gamma)= {1 \over 2} \sum_{\sigma = \pm 1} 
              \{ \sqrt{5/4\pi} d^2_{m \sigma}(\theta_\gamma) \}^2 ,
\label{eq:2.25}
\end{equation}
where $P_m$ denote probabilities in the magnetic substates $m$, and 
$W_m(\theta_\gamma)$ denote $E2$ $\gamma$-ray angular distributions 
of $m$-substates. The intensity distributions, 
$W_m(\theta_\gamma)$ are well-known,\cite{Gamma} 
i.e., $W_0(\theta)= (15/8\pi)\sin^2 \theta \cos^2 \theta$, 
$W_{\pm 1}(\theta)= (5/16\pi)(1-3\cos^2 \theta +4\cos^4 \theta)$ 
and $W_{\pm 2}(\theta)= (5/16\pi)(1- \cos^4 \theta)$. 
In Fig.~2, they are shown for convenience, 
from the top panel $W_0(\theta)$, $W_{\pm 1}(\theta)$ and 
$W_{\pm 2}(\theta)$, respectively.

IReS $4\pi$ $\gamma$-data were given for the probabilities in the 
magnetic substates, where three different quantization axes are taken:  
(a) the beam axis, (b) the axis normal to the scattering plane, 
and (c) the axis perpendicular to (a) and (b) axes.\cite{Nouicer99}
Note that (b) is important for understanding spin alignments, 
where the $z$-axis is normal to the scattering plane. 
Later in Fig.~8, the "m=0" pattern is seen in (b), which suggests 
fragment spins are in the scattering plane. 
In contrast, in Fig.~3, we see the "m=2" pattern clearly in (b), 
where theoretical results of the maximum-aligned configuration 
($J=38$ with $l'=34(I'=4)$) are displayed, in three panels 
(a)$\sim$(c) corresponding to three quantization axes.

We specify the initial beam direction $\boldsymbol{k}$ 
%%in Eqs. (\ref{eq:2.22}) and (\ref{eq:2.24}). 
in the coordinates corresponding to each quantization axis. 
The final fragment directions $\boldsymbol{k}'$ are determined 
by the experimental procedures,\cite{Beck2000, Nouicer99} 
in which two large-area position-sensitive particle detectors are located 
symmetrically on either side of the beam axis in the horizontal plane, 
to take triple coincidence of fragment-fragment-$\gamma$. 
The $^{28}\rm Si$ fragments are detected in the angular range of 
$\theta_{\rm cm} = 90^\circ \pm 7^\circ$ in the reaction plane, 
and with the vertical acceptance $\pm 4^\circ$. Thus the final 
fragment direction is approximately corresponding to (c)-axis.

For quantitative analyses of the angular correlations, 
it is important to take the angular range 
of particle detectors into account.  %%%%%%\cite{Rybicki} 
One may think that the size of the particle detector is enough small 
to neglect the effect. 
But in extremely high spin resonances such as $L \sim 38$, 
amplitudes of the partial waves rapidly oscillate along $\theta_{\rm cm}$ 
in the reaction plane; it needs only $2^\circ$ from the maximum to zero 
amplitude as is seen in Fig.~1.
Hence we need to take the average over the detector area, 
especially along $\theta_{\rm cm}$. 
Note that such an effect is indispensably important for the angular 
correlations in the single excitation channel, 
because due to the Bohr condition we have no $m=\pm 1$ component 
at exact $\theta_{\rm cm} = 90^\circ$,\cite{BohrCond} 
while actually we have rather large $m=\pm 1$ components as seen 
in (a)-axis of the experimental data\cite{Beck2000}    %%[10] 
(see Fig.~11). 
On the other hand, the vertical acceptance $\pm 4^\circ$ is not serious. 
Since the orbital angular momentum vector is almost perpendicular to the 
reaction plane, the waves very slowly vary in the vertical direction, 
though they propagate rapidly oscillating in the reaction plane. 
Therefore we need to integrate over only one dimension, 
i.e., along $\theta_{\rm cm}$ with interval $90^\circ \pm 7^\circ$. 
To be sure, we have compared the results between those with 
one-dimensional integration and with two-dimensional one. 
Actually the differences are very small, less than 1\% of 
the total amounts.

To compare theoretical results with experiment, 
background effects in the angular correlation measurements 
should be considered.\cite{Beck2000}  %%[10] 
Since the excitation yields in the single and mutual $2^+$ channels 
are seen to have backgrounds of over 50\%,\cite{BettsPRL2}    %%[6], 
some contributions are expected, probably from non-resonant 
aligned configurations. 
As is shown in Fig.~3, the aligned configurations appear 
to provide $|m|=2$ contributions in (b), while they provide dominant 
$|m|=1$ contributions together with significant $m=0$ ones in (a) and (c). 
Note that angular correlations with the aligned configuration 
for the single excitation appear to be almost the same as those for 
the mutual excitation shown in Fig.~3.

\subsection{Internal wave functions: molecular model 
of the $^{28}Si + {}^{28}Si$ system}

%
%%%%%%%%%%%%%%%%%%%%%%%%%%%%%%%%%%%%%%%%%%%%%%%%%%%%%%
%%%%%%%%%%%%%%%%%%%%%%%%%%%%%%%%%%%%%%%%%%%%%%%%%%%%%%
\begin{figure}[b]
\centerline{\includegraphics[height=4.0 cm]
                             {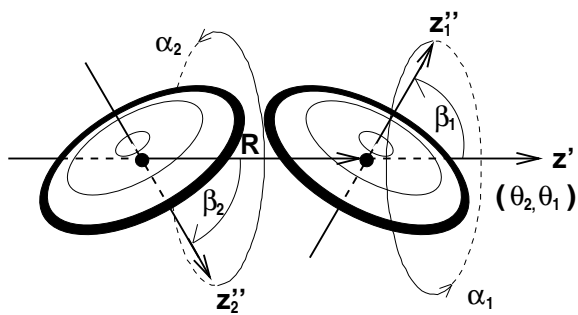}}
\caption{ The dinuclear configuration and the coordinates 
in the rotating molecular frame for an oblate-oblate system. 
The molecular $z'$-axis and the seven degrees of freedom of 
the system are displayed, where the 
$\alpha_1$- and $\alpha_2$-degrees are to be combined 
into $\theta_3 = (\alpha_1 + \alpha_2)/2$ 
and the degree of twisting $\alpha=(\alpha_1 - \alpha_2)/2$. 
The figure is taken from Ref.~\citen{UeNewI}. 
It is the same as published in Refs.~\citen{Ue94} 
and \citen{UeSuppl}. 
}
                 \label{fig:4}
\end{figure}
%%%%%%%%%%%%%%%%%%%%%%%%%%%%%%%%%%%%%%%%%%%%%%%%%%%%%%

Internal wave functions for the $R$-matrix calculations are, 
of course, those of dinuclear molecular model, 
the characteristic features of which are briefly reminded here, 
since they are given in detail in the paper~I.\cite{UeNewI}

Interaction between two nuclei is described 
with internal collective variables, 
i.e., the orientations of the poles of the constituent nuclei 
in the rotating molecular frame.
Assuming a constant deformation 
and the axial symmetry of the constituent nuclei, we are dealing with 
seven degrees of freedom, 
\begin{equation}
(q_{i})= (\theta _{1},\theta _{2},\theta _{3}, R,\alpha,\beta_1,\beta_2),
%
%%\eqno(2.2)PaperI
\label{eq:2.26}
\end{equation}
as illustrated in Fig.~4, 
where $(R, \theta_2, \theta_1)$ is the relative vector of 
the two ${}^{28}\rm Si$.
As internal degrees of freedom, the orientations of the symmetry axes 
of the two ${}^{28}\rm Si$ are described with 
the Euler angles  $(\alpha_i,\beta_i)$ which refer to the molecular axes.
$\alpha_1$ and $\alpha_2$ are combined into 
$\theta_3=(\alpha_1 + \alpha_2)/2$ and $\alpha=(\alpha_1 - \alpha_2)/2$.
All those orientation dependences of the interaction are described 
by a folding-model potential.

It is found that at high spins 
the dinuclear system with oblate-deformed constituent nuclei 
has the equilibrium in an equator-equator touching configuration 
with the parallel principal axes, and that the relative distance 
between the two $^{28}\rm Si$ nuclei is $7 - 8$fm indicating 
a nuclear compound system with hyperdeformation.
The barrier position (or saddle point) is $9 - 10$fm, 
greatly outside from that of the usual optical potential.
Molecular ground state configurations are well bound by the barrier 
up to $J=40$. 
This theoretical maximum spin is in accord with 
the bumps observed in the grazing angular momenta.

Since the interaction potential does not couple the states of 
different $K$-values, at first, 
we assume the eigenstates of the system to be the rotation-vibration 
type with a good $K$-quantum number, 
\begin{equation}
   \Psi_\lambda \sim D_{MK}^J (\theta_i) 
                      \chi_K(R, \alpha,\beta_1,\beta_2), 
%
%%\eqno(2.19)PaperI
\label{eq:2.27}
\end{equation}
then the problem to be solved is of internal motions associated 
with the variables $(R, \alpha, \beta_1, \beta_2)$. 
Couplings among various molecular configurations are 
taken into account by the method of normal mode around the equilibrium 
configuration, which gives rise to the molecular modes of excitation. 
We expand the effective potential at the energy minimum point 
(the equilibrium relative distance $R_{\rm e}=7.6$fm, $\alpha=\pi/2$ 
and $\beta_1=\beta_2=\pi/2$), 
and adopt the harmonic approximation to obtain the normal modes; 
the results are as follows. 
The radial motion has no coupling with the other angle variables, 
and it is an independent mode in itself. 
The motions associated with the $\beta$-degrees are well confined 
to be vibrational ones, 
which are classified into new modes, 
{\it butterfly}: $\beta_+ =( \Delta \beta_1 + \Delta \beta_2 )/ \sqrt 2$ 
and 
{\it anti-butterfly}: $\beta_- =(\Delta \beta_1-\Delta \beta_2)/\sqrt 2$ 
around  $\alpha = \pi/2$, respectively.
As for the $\alpha$-degree, the confinement 
in the present folding potential appears to be unexpectedly weak, 
and hence the motion is close to a free rotation, 
which we call {\it twisting rotational mode}. 
Thus we write the internal wave functions as 
\begin{equation}
\chi_{K}(R,\alpha, \beta_1,\beta_2) 
   = f_n(R)\phi_K(\alpha)\varphi_{n_+}^+(\beta_+ ,\alpha) 
                       \varphi_{n_-}^-(\beta_- ,\alpha) , 
%%\eqno(B.5)PaperI
\label{eq:2.28}
\end{equation}
where $\varphi_{n_+}^+(\beta_+ ,\alpha)$ 
and $\varphi_{n_-}^-(\beta_- ,\alpha)$ 
denote the functions describing the $\beta_+$ and $\beta_-$ modes, 
respectively. 
$\alpha$'s in $\varphi_{n_+}^+(\beta_+ ,\alpha)$ and 
$\varphi_{n_-}^-(\beta_- ,\alpha)$ are due to 
the strengths of the confinements in the $\beta_\pm$-motions, 
which vary together with the $\alpha$-motion.

The eigenenergy of the system is given as 
\begin{eqnarray}
 E^J(n,n_+,n_-,K,(\nu, \pi_\alpha)) 
& = 
& E_0(R_{\rm e})
    +{\hbar^2 \over 2} \bigg[{J(J+1)-K^2 -1 \over \mu R_{\rm e}^2} 
                                        + {K^2 -2 \over 2I} \bigg]  
\nonumber \\
    &  & + \Big( n+{1\over 2} \Big) \hbar \omega_R 
\nonumber \\
    &  &  + \big( n_+  +n_-  + 1 \big)  \hbar \omega_0 
        + E_\nu^\alpha(\pi_\alpha) ,
\label{eq:2.29}
\end{eqnarray}
where the energy is specified by the quantum numbers 
$(n, n_+, n_-, K, (\nu, \pi_\alpha))$, with $\nu$ as a dominant frequency 
of the $\alpha$-motion and with the parity $\pi_\alpha$ 
about the reflection with respect to $\alpha= \pi /2$.
The first and second terms on the r.h.s. of Eq.~(\ref{eq:2.29}) 
are constant energies that are given by the interaction potential 
and the centrifugal energy at the equilibrium, respectively, 
where $I$ denotes the moment of inertia of the constituent nuclei, 
$^{28}\rm Si$, the value of which is estimated 
from the excitation energy of the $2^+_1$ state. 
The term $( n_+  +n_- + 1)\hbar \omega_0$ gives vibrational energies 
for the $\beta$-motions without the $\alpha$-dependence, 
and finally 
$ E_\nu^\alpha(\pi_\alpha)$ is the energy for the $\alpha$-motion 
including the $\alpha$-dependent contribution to the energy of the 
$\beta$-motions. 
The values of the vibrational energy quanta for the butterfly and 
the anti-butterfly modes are both about $4$MeV, but the excitation 
energies of those modes appear twice, $8$MeV, since states of $K=even$ 
with one vibrational quantum are not allowed due to the boson symmetry. 
The excitation energy is close to that of the radial excitation. 
Although the excited state of the radial mode is not bound in the present 
calculations, the possibility of the radial-mode resonance is not 
completely excluded, because it is likely that the interaction between 
two $^{28}\rm Si$ would be more attractive than the present folding 
potential with the frozen density approximation. 
It is noted that the $\alpha$-dependence in 
$\varphi_{n_\pm}^\pm(\beta_\pm ,\alpha)$ is rather tedious in calculations 
of the partial widths in the next section. 
Hence we dropped the $\alpha$-dependence in them, i.e., to be 
$\varphi_{n_\pm}(\beta_\pm)$ with the average value $4$MeV of 
$\hbar \omega_\beta$, 
which means that the eigenenergies of the butterfly and anti-butterfly 
modes are taken to be degenerate. However the properties of those modes 
are essentially retained with no problem. 
Of course, for exact calculations, we need total wave functions with 
the parity and boson symmetries, which are described in Appendix B 
of the paper~I,\cite{UeNewI} some examples of the wave functions 
of the normal modes being also given there in Appendix C.

\vskip 5 true mm

%%%%%%%%%%%%%%%%%%%%%%%%%%%%%%%%%%%%%%%%%%%%%%%%%%%%%%
%%%%%%%%%%%\begin{figure}[b]
\begin{wrapfigure}{r}{6.6cm}
\centerline{\includegraphics[height=5.4 cm]
                             {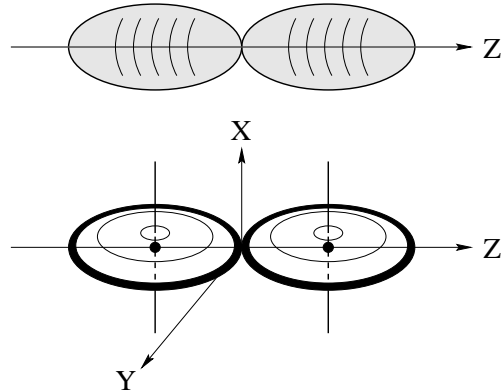}}
\caption{
Equilibrium configurations of two di-nuclear systems. 
$\,$ The upper portion 
is for $^{24}\rm Mg-^{24}Mg$ 
$\,$ and $\,$ the lower one for $^{28}\rm Si-^{28}Si$.
The figure is taken from Ref.~\citen{UeNewI}.
}
                         \label{fig:5}
\end{wrapfigure}
%%%%%%%%%\end{figure}
%%%%%%%%%%%%%%%%%%%%%%%%%%%%%%%%%%%%%%%%%%%%%%%%%%%%%%

\noindent
{\it Wobbling motion ($K$-mixed states)}

One of the characteristic features of the spectrum 
obtained theoretically is a series of low-energy 
$K$-rotational excitation due to axial asymmetry around molecular z-axis, 
which is in contrast with the $^{24}\rm Mg+{}^{24}Mg$ case.\cite{Ue89,Ue93}
One can understand the reason immediately from Fig.~5, 
where the upper configuration($^{24}\rm Mg+{}^{24}Mg$) has axial symmetry 
as a total system, but the lower one for $^{28}\rm {Si} +{}^{28}\rm {Si}$ 
has axial asymmetry.

A triaxial system preferentially rotates around the axis 
with the largest moment of inertia.  
By the definition of the axes in the lower panel of Fig.~5, 
we have the moments of inertia as $I_X > I_Y >> I_Z$, 
due to the nuclear shape.
Thus the system, which is seen as two pancake-like 
objects ($^{28}\rm Si$'s) touching side-by-side, 
rotates around $X$-axis normal to the reaction plane.  
Such a motion is called wobbling, and 
$K$ is not a good quantum number. 
Namely, we expect that the eigenstates are $K$-mixed.

%%%%%%%%%%%%%%%%%%%%%%%%%%%%%%%%%%%%%%%%%%%%%%%%%%%%%%%%%%%%%%%%%%%%
%%%%%%%%%%%%%%%%%%%%%%%%%%%%%%%%%%%%%%%%%%%%%%%%%%%%%%%%%%%%%%%%%%%%
\begin{figure}[t]
\centerline{\includegraphics[height=7.6 cm]
                             {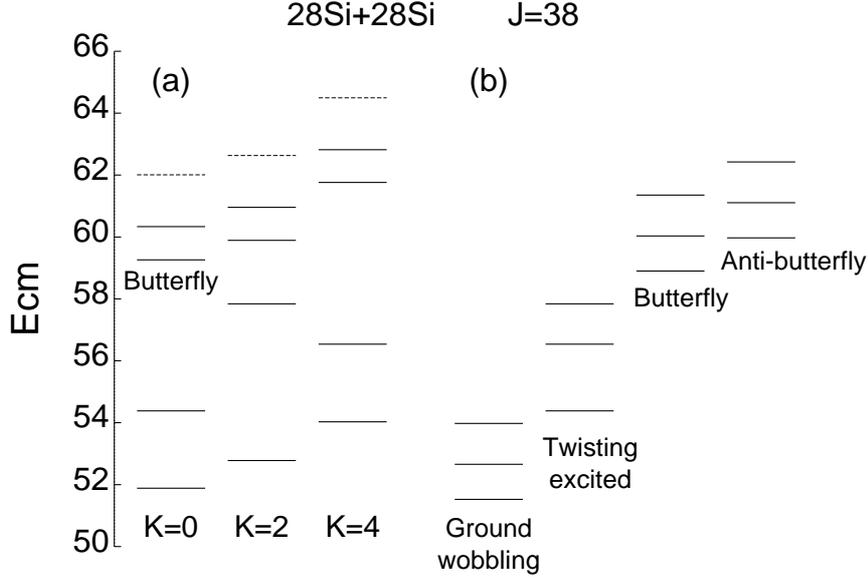}}
\caption{
Energy spectra of the $^{28}\rm Si+{}^{28}Si$ system for $J=38$. 
(a) Molecular normal modes without $K$-mixing. 
(b) After $K$-mixing, with indications of the modes under the levels. 
The figure is taken from Ref.~\citen{UeNewI}.
}
                        \label{fig:6}
\end{figure}
%%%%%%%%%%%%%%%%%%%%%%%%%%%%%%%%%%%%%%%%%%%%%%%%%%%%%%%%%%%%%%%%%%%%

Regarding the $^{28}\rm {Si} +{}^{28}\rm {Si}$ system as a
triaxial rotator, as it is usual for polyatomic molecules, 
we diagonalize the hamiltonian with an inertia tensor of the axial 
asymmetry, which gives rise to mixings of $K$-projections of the 
total spin $J$.\cite{BohrTEXT2} 
The resultant motion should be called as "wobbling mode".\cite{Wobb} 
The energy spectrum is displayed in Fig.~6(b), compared with 
the spectrum without $K$-mixing in Fig.~6(a).
Now the states of low lying $K$-series are not the eigenstates 
by themselves, but are recomposed into new states.  
It is much interesting that we again obtain several states including 
the $K=0$ component as a result of $K$-mixing, 
which should show up themselves in the scattering.
Those states are closely located in energy and so in good agreement 
with several fine peaks observed in the experiment.

As an analytical prescription, in the high spin limit 
($K/J  \sim 0$), the solution is a gaussian, or a gaussian 
multiplied by an Hermite polynomial, 
\begin{equation}
  F_n(K) = H_n \left({ K \over b } \right)
  \exp \biggl[-{1 \over 2} \biggl( { K \over b } \biggr)^2 \biggr] .
%
%%\eqno(4.7)PaperI
\label{eq:2.30}
\end{equation} 
The width parameter $b$ is given by 
\begin{equation}
b= (2J^2  {I_K / \Delta } )^{1/4},
%
%%\eqno(4.8)PaperI
\label{eq:2.31}
\end{equation} 
which depends on the ratio between the 
non-axiality (the difference between $I_X$ and $I_Y$) and $I_Z$ as 
\begin{equation}
{1 \over \Delta}= {1 \over 2}\bigg({1 \over I_Y} -{1 \over I_{X}}\bigg), 
\qquad
{ 1 \over I_K} =  {1 \over I_Z} 
            - {1 \over 2}\bigg({1 \over I_Y} +{1 \over I_{X}}\bigg). 
%
%%\eqno(4.4)(4.5)PaperI
\label{eq:2.32} 
\end{equation}

To calculate angular correlations we use those analytic forms 
in Eq.~(\ref{eq:2.30}), which is simple and intuitive way to understand 
the extent of $K$-mixing.
Of course we can utilize numerical values obtained 
in the diagonalization procedure, but the values are almost the same 
as those given by the analytic form.
For the lowest state $F_0 (K)$ of Eq.~(\ref{eq:2.30}), 
we have the wave function for the wobbling ground state as 
\begin{equation}
\Psi^{JM}_\lambda \sim \sum_K  \exp (-K^2 / 2b^2)  D_{MK}^J (\theta_i) 
                      \chi_{K}(R, \alpha,\beta_1,\beta_2) , 
%
%%\eqno(4.11)PaperI
\label{eq:2.33}
\end{equation} 
where it should be emphasized that, in general, $\chi_K$ can be 
any molecular mode of triaxial deformations, such as the ground-state 
configuration (parallel equator-equator one), 
the butterfly mode and the anti-butterfly mode. 
Then the spin orientations of two $^{28}\rm Si$ nuclei are expected 
to be in the plane, consistent with $"m=0"$, because the nuclei 
rotate around the axes perpendicular 
to their symmetry ones. 
The magnitude of $b$ estimated by Eq.~(\ref{eq:2.31}) is $1.85$, 
for example, for the values of the moments of inertia used 
in the calculations for the energy spectrum in Fig.~6. 
This is the largest value expected, because we assumed a static 
configuration there, in which the zero-point motions of the twisting 
and butterfly modes are neglected. 
Thus, a value of $b=1.5$ is adopted for the calculations of the 
physical quantities in the next section. 
There, it is also examined 
to what extent a change of $b$ affects the spin disalignments 
with the orbital angular momentum.

As for the relation between the molecular model and the asymmetric 
rotator intuitively introduced, we mention that it is clarified 
in \S4.2   %%%%subsection 4.2. 
of the paper~I\,\cite{UeNewI}  
by using a simple example of the dinuclear system of "one deformed 
nucleus and a spherical nucleus", and thus we do not repeat here.

\section{Results}

\subsection{Structures and partial decay widths of the resonance states}

\begin{table}[b]
\caption{Theoretical estimates for partial widths (in keV) in the elastic, 
single $2^+$ and mutual $2^+$ channels of the molecular resonances 
with spin $J=38$ for the theoretical level energies, 
which are obtained in \S2.3.     %%%%subsection 2.3. 
For the wobbling strength, $b=1.5$ is adopted. 
In parenthesis, widths are given with the assumption of 
the resonance energy $55.8$MeV.
      }\label{table:2}
      \begin{center}
      \begin{tabular}{ccccc} \hline \hline
Molecular states & Energy
      &  $\Gamma_{el}$ &  $\Gamma_{2^+}$ &  $\Gamma_{(2^+,2^+)}$
\\ \hline 
molecular ground state  & 51.5MeV &{2.1}&{1.2}&{0.16} \\
{}                  &   &{(19)}&{(13)}&{(2.2)} \\
twisting & 54.4MeV      &{0.36}&{0.24}&{6.2} \\ 
{}                  &   &{(0.69)}&{(0.46)}&{(13)} \\
butterfly & 58.9MeV     &{28}&{1.5}&{21} \\
{}                  &   &{(8.3)}&{(0.43)}&{(5.0)} \\
anti-butterfly & 60.3MeV  &{40}&{1.5}&{69} \\
{}                  &     &{(8.3)}&{(0.25)}&{(15)} \\ \hline
\end{tabular}
\end{center}
\end{table}

In order to analyze structures of the resonance states by the molecular 
model, we have estimated their partial widths.  
In Table \RMNa , the partial widths  %%%of the states 
of the molecular normal modes with spin $J=38$ are given, 
up to the mutual $2^+$ channel. 
The results for the $4^+$ channels are not shown, 
since the experimental angular distributions of those channels do not 
exhibit resonance behaviors.\cite{Beck2000,Nouicer99} 
Magnitudes of the widths are calculated with the theoretical level energies, 
and are obtained to be in the range of several keV to several tens keV, 
which is consistent with experiment. 
For comparison, the magnitudes of the widths are also calculated 
with the assumption that the energy of each state is shifted to 
the observed resonance energy of $55.8$MeV, 
and are given in parenthesis. 
Among the elastic and inelastic channels, characteristic features 
are seen for each normal mode. 
As for the molecular ground state, 
the elastic width is larger than those in the excited channels, 
which is inconsistent with the experimental characteristics 
of the narrow resonances. Those weak excitations of 
the theoretical prediction are inferred to be due to 
the weak confinement in the $\alpha$-degree of freedom, which permits 
almost free rotation for the solution with the folding potential. 
In the butterfly and twisting modes, the mutual $2^+$ channel shows 
strong excitation, but the single $2^+$ excitation is very weak, 
and thus the characteristics are inconsistent with experiment. 
Therefore neither of the normal modes is completely 
consistent with the characteristics of the enhanced excitations 
seen at $E_{\rm cm}=55.8$MeV.

It is, however, meaningful to investigate with a stronger confinement 
in the $\alpha$-degree of freedom for the molecular ground state, 
because, in the touching configuration, the confinement possibly 
increases due to induced deformations of the constituent nuclei. 
So we investigate the cases with stronger confinements 
for the $\alpha$-degree with the harmonic oscillator, 
by using $\hbar \omega_\alpha$ as a parameter. 
This means that we introduce a stronger confinement of the gaussian 
distribution around the equilibrium, instead of the 
internal-rotation-like motions, although 
how much the strength of the confinement should be increased cannot 
be predicted by the folding model within the frozen density approximation, 
but is to be quantitatively investigated  with the polarization effects 
or by dynamical treatments of deformations of the constituent ions.
The polarization effects of the ions expected to appear will be discussed 
later in \S4.3. 

\begin{table}[h]
\caption{Theoretical estimates for partial widths (in keV) 
in the elastic, single $2^+$ and mutual $2^+$ channels 
for the molecular ground state with spin $J=38$ 
at the experimental resonance energy $E_{\rm cm}=55.8$MeV.
Rather confined configurations for the molecular ground state 
are inspected for 
(case A: $\hbar \omega_\alpha =2$MeV, $\hbar \omega_\beta =4$MeV), 
(case B: $\hbar \omega_\alpha = \hbar \omega_\beta =4$MeV) 
and (case C: $\hbar \omega_\alpha = \hbar \omega_\beta =12$MeV). 
For the wobbling strength, $b=1.5$ is adopted.   
      }\label{table:3}
      \begin{center}
      \begin{tabular}{cccc} \hline \hline
 States  &  $\Gamma_{el}$ &  $\Gamma_{2^+}$ &  $\Gamma_{(2^+,2^+)}$
 \\ \hline     
molecular ground state A  & 17  & 12  & 2.9 \\ 
molecular ground state B  & 12  & 9.2 & 4.3 \\ 
molecular ground state C  & 2.6 & 3.5 & 3.0 \\  \hline
\end{tabular}
\end{center}
\end{table}
%
%
%%%%%%* for calcu. b=1.5,(sigK=b**2/2=1.125), $Re=7.6$fm.
%%%%%%A: $\hbar \omega_\alpha =2$MeV
%%%%%%B: $\hbar \omega_\alpha =4$MeV
%%%%%%C: $\hbar \omega_\beta =12$MeV, $\hbar \omega_\alpha =12$MeV
%
%% cf.  a_alpha=1.6(CLUSTER99) -> hw=0.475MeV
%%      b=1.5(a_theta=b^2=2.25): a_alpha=2.25 -> hw=0.67MeV
%

In Table \RMNb , a couple of configurations with stronger confinements 
are introduced, and the estimations of 
the partial widths are given at the energy $55.8$MeV. 
For the configurations named A, B and C, strengths of the confinements 
are set to be $\hbar \omega_\alpha =2$MeV, $4$MeV and $12$MeV, 
respectively, where in the last case $\hbar \omega_\beta =12$MeV is 
taken consistently with the large value $\hbar \omega_\alpha =12$MeV. 
As the confinement increases, the magnitude of the partial 
widths in the elastic channel decreases, while excitations 
to the single and mutual channels are relatively enhanced. 
With the configuration C, 
characteristics of the partial widths are in good agreement 
with the experimental data.\cite{BettsPRL2} 
Moreover the value of the elastic partial width $2.6$keV is 
enough small to reproduce the experimentally suggested value, 
"a few keV".\cite{BettsPRL2} 
Thus the molecular ground state with a well confined configuration 
is a candidate for the narrow resonance at $55.8$MeV, which is 
corroborated by the following analyses on fragment angular 
distributions and fragment-fragment-$\gamma$ angular correlations.

\subsection{Fragment angular distributions for the resonance 
            at $E_{\rm cm}=55.8$MeV}

%% FFAD %%%%%%%%%%%%%%%%%%%%%%%%%%%%%%%%%%%%%%%%%%%%%%%%%%%%%%%%%%%%%%%%

\begin{figure}[b]  % FIG-7(FFAD) one set
\centerline{\includegraphics[height=10.8 cm]
%%%%%%%%%%%%\centerline{\includegraphics[height=9.8 cm]
                             {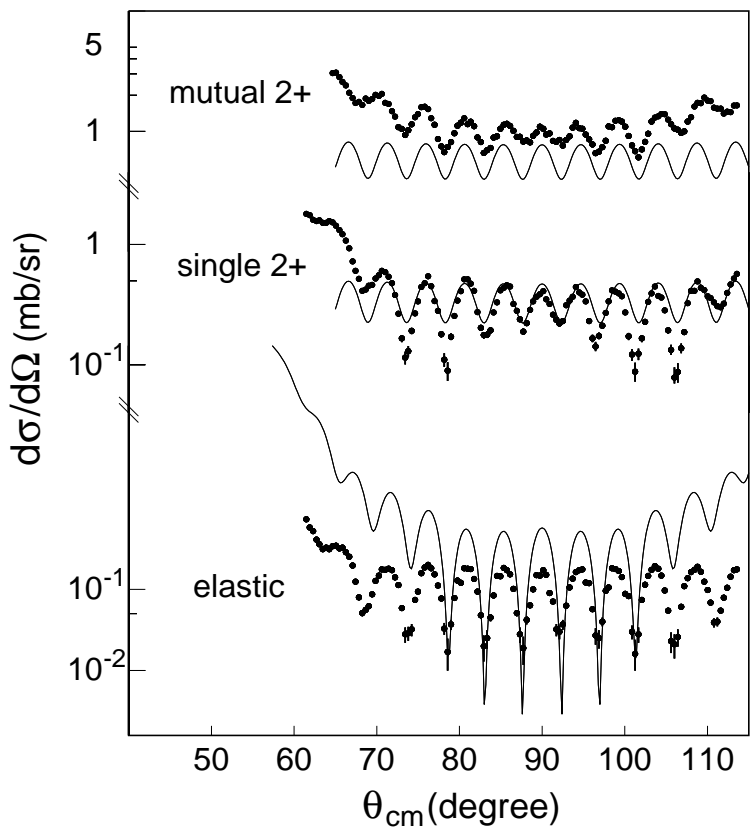}}
\caption{Fragment angular distributions 
for the elastic and inelastic scattering 
for $^{28}\rm Si+{}^{28}Si$ at $E_{\rm cm}=55.8$MeV. 
From the bottom those for the elastic, single $2^+$ and mutual $2^+$ 
are shown.
Solid curves are for theoretical results of the molecular ground state 
with wobbling motion (configuration C), compared with 
the experimental data.\cite{Nouicer99}
}
                      \label{fig:7}
\end{figure}
%%%%%%%%%%%%%%%%%%%%%%%%%%%%%%%%%%%%%%%%%%%%%%%%%%%%%%%%%%%%%%%%%%%%%%%%

In Fig.~7, fragment angular distributions are displayed 
for the elastic scattering, 
the single $2^+$ and the mutual $2^+$ excitations.
Theoretical angular distributions with the configuration C 
of the wobbling ground state are compared with the recent 
data,\cite{Nouicer99}  %%[16],
where their magnitudes are normalized to the single excitation.
For the excitations, constant backgrounds of 40\% are assumed, 
because the background yields of about 50\% are seen in the 
experimental data observed by Betts et al.\cite{BettsPRL2}  %%[6].
In the elastic scattering, by the phenomenological use 
of the potential scattering term as in Eq.~(\ref{eq:2.16}), 
calculations are made with strong absorption of 
lower partial waves than the grazing one to reproduce rapid 
decrease in the differential cross sections toward 
$\theta_{\rm cm}=90^{\circ}$.\cite{BettsC-a, BettsC-b}  %%[1,2].
Such a rapid decrease is usually seen 
in the heavy-ion scattering,\cite{Frahn, BMunzinger}  %%[25, 32],
and it would be one of necessary conditions for the observation of 
the quite narrow resonances in the elastic scattering. 
We see good fits for the oscillations, the period of which is 
close to $L=38$ Legendre polynomials. Magnitudes of the cross sections 
are qualitatively in good agreement with the data, although the 
strengths of the excitations are still slightly weak. 

\subsection{Angular correlations}

%%%%%%%%%%%%%%%%%%%%%%%%%%%%%%%%%%%%%%%%%%%%%%%%%%%%%%%%%%%%%%%%%%%%%%%
\begin{figure}[b]    % FIG-8(Ang.-Corr. Mol. Ground State) one set
\centerline{\includegraphics[height=10.5 cm]           %%%%9.6 cm]
                             {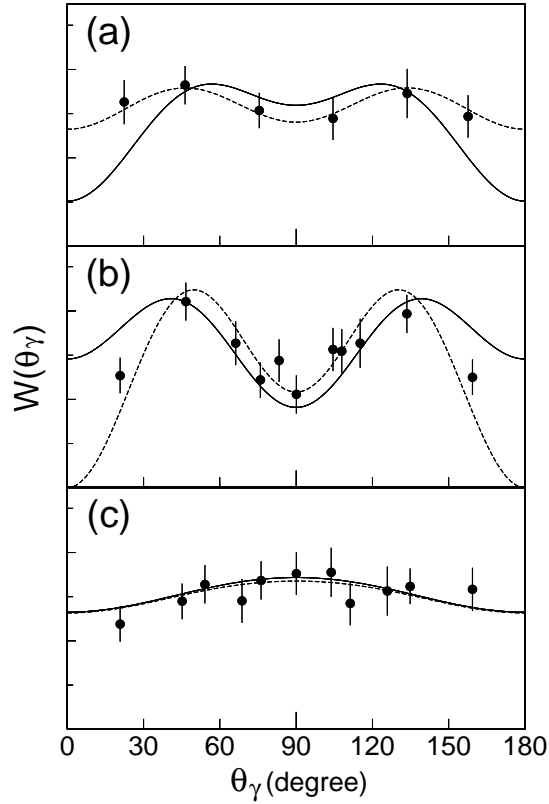}}
\caption{$\gamma$-ray intensities for 
particle-particle-$\gamma$ angular correlations 
in the mutual $2^+$ channel.
Experimental data are given by dots, 
and $\chi^2$-fits curves are displayed by dashed lines. 
The theoretical results obtained with the molecular ground state 
(configuration C: $\hbar \omega_\alpha =\hbar \omega_\beta =12$MeV) 
are shown by solid lines. 
For the quantization axes in three panels, see text. 
}
\label{fig:8}    %%% FIG-8 END
\end{figure} 
%%%%%%%%%%%%%%%%%%%%%%%%%%%%%%%%%%%%%%%%%%%%%%%%%%%%%%%%%%%%%%%%%%%

In the decay process from the molecular resonance state, 
$^{28}\rm Si$-fragments emit $\gamma$-rays due to the transition from 
the first excited state to the ground state. 
In the experiment, particle detectors are set to catch the fragments 
in the direction 
perpendicular to the incident beam, and the coincident $\gamma$ rays 
are detected by the $4\pi$ system of Eurogam Phase \RMNb .$\,\,$
Figure~8 displays the angular correlation data,\cite{Nouicer99}  %%[16] 
i.e., $\gamma$-ray intensity distributions of the $E2$-transition 
observed in the mutual $2^+$ channel and $\chi^2$-fits shown by 
dashed lines. Three different quantization axes are taken in panels 
(a)$-$(c), respectively: 
(a) beam direction, 
(b) $z$-axis normal to the scattering plane, and 
(c) $z$-axis perpendicular to those of (a) and (b). 
Since $^{28}\rm Si$ fragments are detected in the angular range of 
$\theta_{\rm cm} = 90^\circ \pm 7^\circ$ in the reaction plane, the 
$z$-axis of (c) corresponds approximately to final fragment directions.

Theoretical results for the molecular ground state (configuration C 
of a strong confinement) are displayed by solid lines. 
In panel (b), we see a typical "m=0" pattern, which means that the spin 
directions are in the reaction plane.
A weak "m=0" pattern is seen also in panel (a), while in panel (c), 
a slight swelling around the center suggests weak "m=2".

\begin{table}[t]
\caption{The experimental and theoretical probabilities $P_m$ 
in the magnetic substates $m$, for the mutual $2^+$ channel of 
the resonance state with spin $J=38$ at $E_{\rm cm}=55.8$MeV. 
The experimental values of $P_m$ are obtained 
by $\chi^2$-fits of the data.\cite{Nouicer99} %%[16].
Theoretical results are given for the molecular ground state, 
where the first line in (a) shows those for the weak confinement 
(solution with the folding potential), while the second line 
in (a) does those for the strong confinement 
(configuration C: $\hbar\omega_\beta = \hbar\omega_\alpha =12$MeV), 
and in (b) and (c) as well. 
For the wobbling strength, $b=1.5$ is adopted. 
      }\label{table:4}
      \begin{center}
      \begin{tabular}{ccccccc} \hline \hline
Quantization   & 
 \multicolumn{2}{c}{$P_{m=0}$} &
 \multicolumn{2}{c}{$P_{m=\pm 1}$} &
 \multicolumn{2}{c}{$P_{m=\pm 2}$} \\
axis     & exp. & theor. & exp. & theor. & exp. & theor. \\ \hline 
(a)&${0.30\pm 0.08}$&${0.51}$ &${0.16\pm 0.04}$&${0.07}$ &${0.18\pm 0.05}$&${0.17}$ \\
   & {}             &${0.20}$ & {}             &${0.06}$ & {}             &${0.34}$ \\
(b)&${0.46\pm 0.05}$&${0.70}$ &${ 0          }$&${0.01}$ &${0.27\pm 0.02}$&${0.14}$ \\
   & {}             &${0.54}$ & {            } &${0.18}$ & {}             &${0.05}$ \\
(c)&${0.14\pm 0.05}$&${0.05}$ &${0.17\pm 0.03}$&${0.08}$ &${0.26\pm 0.04}$&${0.40}$ \\
   & {}             &${0.14}$ & {}             &${0.17}$ & {}             &${0.26}$ 
\\ \hline 
\end{tabular}
\end{center}
\end{table}

%%%%%%%%%%%%%%%%%%%%%%%%%%%%%%%%%%%%%%%%%%%%%%%%%%%%%%%%%%%%%%
%%%%%%%%%%%%%%%%%%% FIG-9, FIG-10(Mutual twist, butt) one set
\begin{figure}[b] 
  \parbox[t]{\halftext}{%   %\def\halftext{.471\textwidth}
  \centerline{\includegraphics[%%width=WIDTH cm,
                             height=11.0 cm]        %%%9.6 cm]
                             {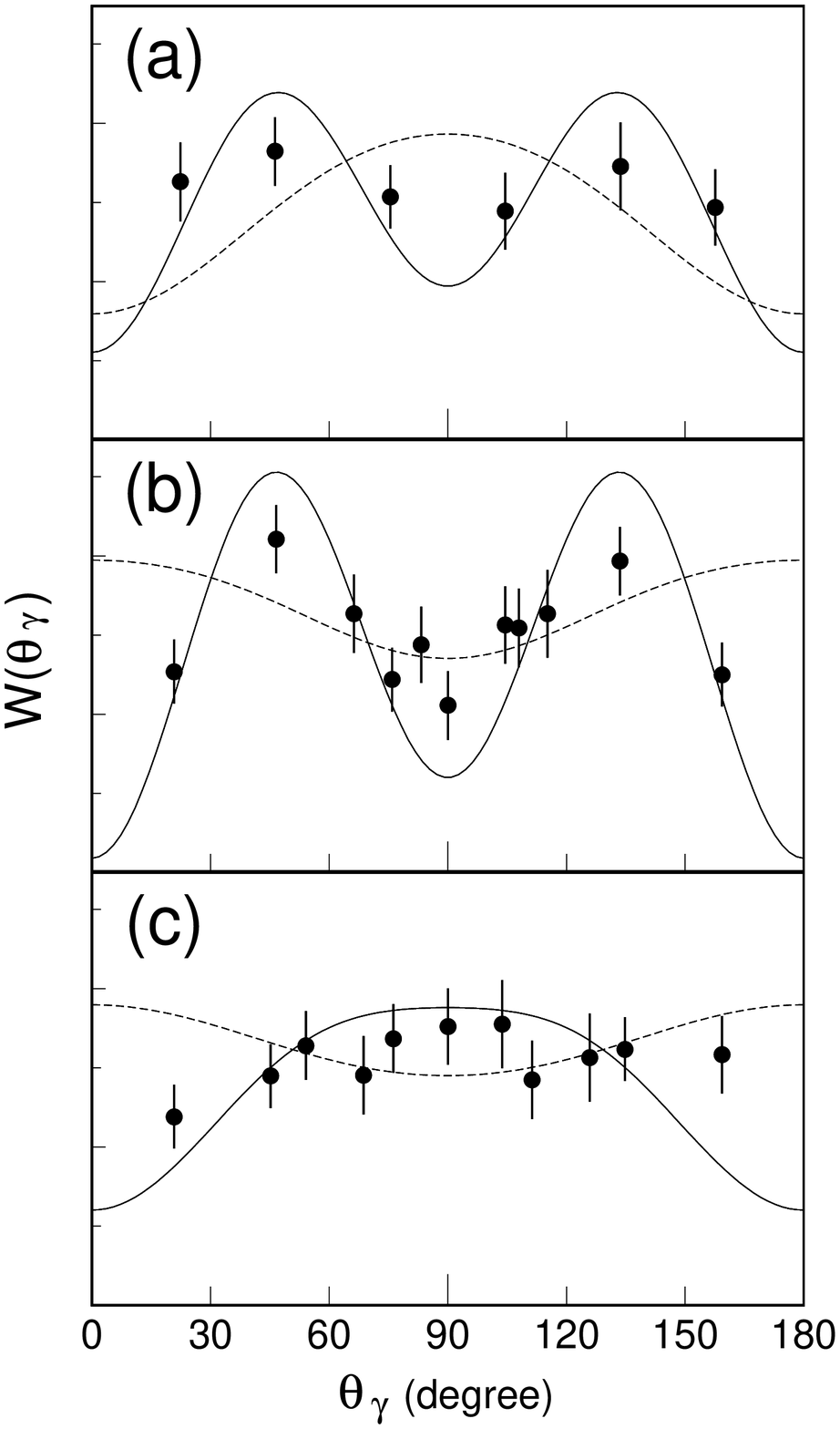}}
\caption{   $\gamma$-ray intensities for particle-particle-$\gamma$ 
angular correlations in the mutual $2^+$ channel.
Experimental data are given by dots. 
Dashed lines show theoretical results of the twisting rotational 
mode, with solid lines for the molecular ground state for comparison. 
For the quantization axes in three panels, see text. }
}
  \label{fig:9}
  \hfill
  \parbox[t]{\halftext}{%   %\def\halftext{.471\textwidth}
  \centerline{\includegraphics[%%width=WIDTH cm,
                             height=11.0 cm]        %%%9.6 cm]
                             {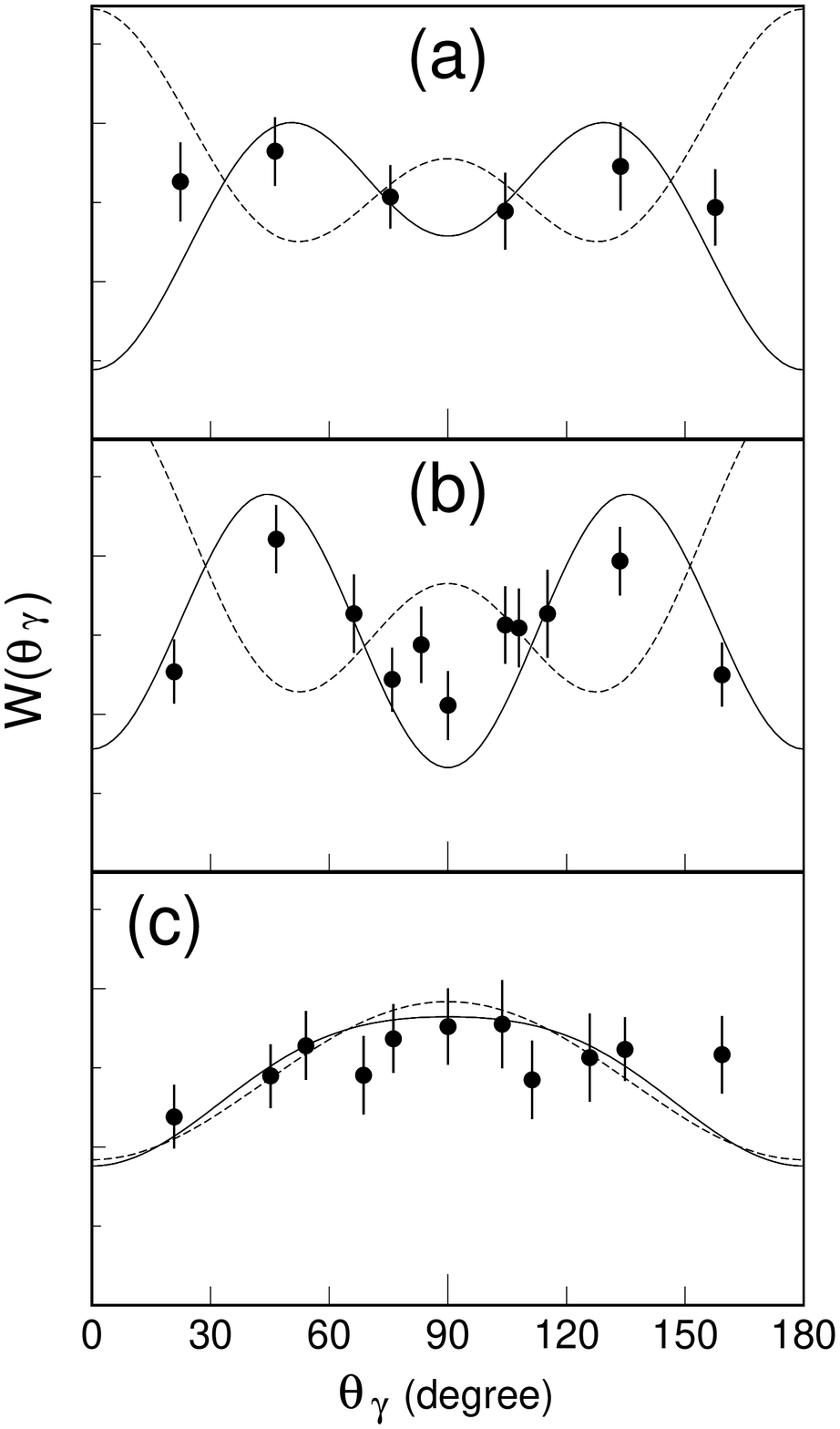}}
\caption{$\gamma$-ray intensities for particle-particle-$\gamma$ 
angular correlations in the mutual $2^+$ channel.
Experimental data are given by dots.
Dashed lines show the theoretical results 
of the butterfly state with $K$-mixing, 
with solid lines for the molecular ground state for comparison. 
For the quantization axes in three panels, see text. } 
}
  \label{fig:10}
\end{figure}    %%% FIG-9, FIG-10 END
%
%%%%%%%%%%%%%%%%%%%%%%%%%%%%%%%%%%%%%%%%%%%%%%%%%%%%%%%%%%%%%%%%
%
%%TABLE&FIGS
%
%%\vfill
%%\clearpage
%
%%\noindent

Probabilities $P_m$ in each magnetic substate $m$ are listed 
in Table \RMNc . 
Theoretical results of the molecular ground state are given 
and compared with the experimental data,\cite{Nouicer99} 
where the upper line in each axis case is for a weak confinement case 
(solution with the folding potential), 
while the second line is for a strong confinement (configuration C), 
respectively. 
In (b), theoretical results reproduce dominant "m=0" characteristics, 
for both configurations of the weak and strong confinements. 
Thus, the "m=0" characteristics of the angular correlations do not 
depend essentially upon the strength of the confinement 
in the $\alpha$-degree of freedom. 
$|m|=2$ contributions in (b) appear slightly weak, compared with 
the experimental data, but this suggests $|m|=2$ 
contribution from the background.\cite{Beck2000} 
As for the values of $P_m$ in (a) and (c), they are also in good 
agreements with the data, although they vary depending 
on the strengths of the confinements.

To clarify properties of the normal mode excitations in the angular 
correlations, we make analyses for the twisting and butterfly modes, 
which are of interest as they exhibit strong excitation to the mutual 
$2^+$ channel.
In Fig.~9, the angular correlations of the twisting mode are displayed 
with dashed lines, and comparison with the molecular ground state 
(solid lines) is given. 
Apparently, the twisting excitation does not fit the experimental data. 
The characteristic feature of twisting is "m=2" dominance in (a), 
which is due to the excited rotational motion in the $\alpha$-degree 
of freedom.

In Fig.~10, results are displayed with dashed lines 
for the butterfly mode with $K$-mixing. 
The results for the anti-butterfly mode appear to be quite 
similar to those of the butterfly mode (not shown here).
Again, solid lines are those for the molecular ground state 
for comparison.
The butterfly excitations show their own characteristics with dominant 
"m=1" patterns in the panels (a) and (b). 
In addition, they exhibit "m=2" dominance in (c), 
due to the butterfly vibrational motions.  
Thus all those theoretical results for the normal mode excitations 
are much different from the data at the resonance energy $55.8$MeV, 
and moreover they show quite distinguishing characteristics.

Note that the configurations of the molecular ground state 
in Figs.~8, 9 and Fig.~10 are not completely the same due to changes 
of the confinement parameter value, and one can confirm differences 
between the results with different strengths of confinements in 
the $\alpha$-degree of freedom. 
In Fig.~9, the most weak confinement is taken from the solution 
with the folding potential, and in Fig.~10 a slightly stronger 
confinement is taken (the configuration A: 
$\hbar\omega_\alpha =2$MeV, $\hbar\omega_\beta =4$MeV). 
All those configurations give "m=0" characteristics, and the 
differences between them are not significant. 

\vfill
\clearpage

%%%%%%%%%%%%%%%%%%%%%%%%%%%%%%%%%%%%%%%%%%%%%%%%%%%%%%%%%%%%%%%%%%%%%%%
%%% Ang. Correlations SINGLE      %%% FIG-11(single chan.) one set 
%
\begin{wrapfigure}{r}{6.6cm}       
%%  \figurebox{60mm}{9cm}
  \centerline{\includegraphics[%%width=WIDTH cm,
                             height=11.0 cm]        %%%9.6 cm]
                             {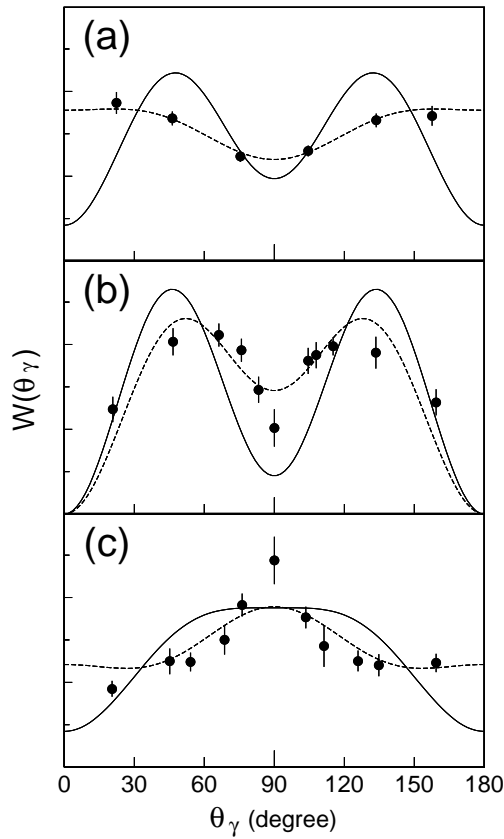}}
\caption{$\gamma$-ray intensities for 
particle-particle-$\gamma$ angular correlations 
in the single $2^+$ channel.
Experimental data are given by dots, 
and $\chi^2$-fits by dashed lines, respectively. 
Solid lines show the theoretical results 
obtained with the molecular ground state with wobbling motion 
(configuration C).
For the quantization axes in three panels, see text. 
}
  \label{fig:11}
\end{wrapfigure}       %%% FIG-11 END
%
%%%%%%%%%%%%%%%%%%%%%%%%%%%%%%%%%%%%%%%%%%%%%%%%%%%%%%%%%%%%%%%%%%%%%%%%

The magnitude of wobbling ($K$-mixing) is taken to be $b=1.5$, 
which is consistent with the asymmetry of the 
${}^{28}\rm {Si} +{}^{28}\rm {Si}$ molecular ground-state 
configuration, as it is estimated to be $b=1.85$ in the limit 
of the strong confinement (without quantum fluctuations). 
The parameter for the $K$-mixing is examined in the range of 
$b=1.2 - 2.0$ with the configuration C. 
The value of $P_{m=0}$ in (b) varies from 0.4 to 0.8 with the 
increasing value of $b$, but for the configurations with the 
strong confinements the resultant curves are seen to be essentially 
the same in those values of $b$.

Next we move to the single $2^+$ excitation. 
In Fig.~11, the experimental data are displayed, together with 
$\chi^2$-fits given by dashed lines.\cite{Beck2000} 
The theoretical results for the molecular ground state 
(configuration C of a strong confinement) are shown 
by solid lines. 
Surprisingly, the experimental data are seen to be quite similar 
to those of the mutual $2^+$ channel, 
and again "m=0" characteristics in (b) are firstly of interest. 
The present model reproduces the characteristics well.
Theoretical results exhibit totally good fits with the data, 
although the fits in panel (a) are not very good, 
with rather large $m=0$ contribution. 
However, contributions of the background from aligned 
configurations are expected to provide $|m|=1$ components 
in (a) and (c), 
and thus those slight deviations of the fits are not serious.

\section{Discussion}

\subsection{Structure of the $^{28}Si - {}^{28}Si$  molecule}

In the paper~I,\cite{UeNewI}  interaction between two $^{28}\rm Si$ nuclei 
is described with the internal collective variables, 
i.e., with the orientations of the poles of the constituent nuclei 
in the rotating molecular frame.
For the dinuclear system with oblate-deformed constituent nuclei, 
an equator-equator touching configuration is found to be the equilibrium 
at high spins where the principal axes of the constituent nuclei are 
parallel. The relative distance between the two $^{28}\rm Si$ nuclei is 
$7 - 8$fm indicating a nuclear compound system with hyperdeformation. 
The barrier position is $9 - 10$fm, 
greatly outside from that of usual optical potentials.
Molecular ground state configurations are well stable by the barrier 
up to $J=40$. 
This theoretical maximum spin is in accord with 
the bumps observed in grazing angular momenta.

Couplings among various molecular configurations 
are taken into account by the method of normal mode around the equilibrium 
configuration, which gives rise to the molecular modes of excitation,
such as the radial vibration, 
the butterfly motion, the anti-butterfly motion and so on. 
The twisting mode ($\nu =4$) appears to be the lowest excitation, 
but the energy may be higher than the present results due to a stronger 
confinement as physically expected, which will be discussed later 
in \S4.3.

Since the equilibrium configuration has a triaxial shape, 
we extend our molecular model so as to include couplings 
between states with different $K$-quantum numbers.  
In practice, we use the asymmetric rotator as an intuitive model, 
which preferentially rotates around the axis of 
the largest moment of inertia, accompanied with $K$-mixing, 
and thus gives rise to the wobbling motion.

\subsection{On the spin disalignments} 

An important physical quantity which probes the structure of 
the resonance states is spin alignments of the outgoing particles. 
The recent experiment on the $^{28}\rm Si+ {}^{28}Si$ resonance at 
$E_{\rm cm}=55.8$MeV suggests $L=J=38$ dominance in the inelastic 
scattering both to the single and mutual $2^+$ excitations.
Measured particle-particle-$\gamma$ correlations show "m=0" dominance 
in $\gamma$-rays.  All these suggest that the spin vectors are 
in the reaction plane.\cite{Beck2000, Nouicer99}

The analyses are made for the molecular model wave functions. 
In the state with the lowest energy of a given angular momentum $J$,  
due to the triaxial configuration, the whole system rotates about 
the axis normal to the reaction plane defined by the two pancake-like 
nuclei. The spins of the $^{28}$Si fragments are thus in this plane, 
since no rotation can occur about the symmetry axes of the $^{28}$Si 
nuclei. Such a property is in agreement with the lack of strong
alignments observed in the fragment angular distributions. 
Actually, "m=0" dominances in the angular correlations 
are well reproduced for the single and mutual $2^+$ excitations.

Characteristic features of the angular correlations in the normal mode 
excitations are found to be dominant "m=2" patterns, i.e., 
for the twisting rotational mode "m=2" appears in (a)-axis, and 
for the butterfly mode it appears in (c)-axis, respectively. 
This means that spin vectors are parallel to the beam direction 
in the twisting mode, while parallel to the fragment direction 
in the butterfly mode.  (The reason for those orientations of the 
spin vectors is explained later.)
One may expect that the twisting mode and/or the butterfly mode 
are favorable for the "m=0" characteristic, 
because spin orientations are in the reaction plane. However, 
more precisely, "m=0" means {\em symmetric around the normal axis}, 
which is satisfied by neither of excited states such as twisting nor
butterfly with a well-defined direction of the spin in the reaction plane. 
The existence of dominant $m=2$ substates, together with "m=1" patterns 
in the other panels, are clear characteristics of the normal modes. 
Thus, it is possible to distinguish among the molecular ground state, 
the twisting excitation and the butterfly excitations, respectively.

Finally, we discuss the observed spin directions in space, 
with respect to the orientation of the molecular $z'$-axis at the time 
of decay, which is related to the positions of the particle detectors 
in the angular correlation measurements. 
For example, in the twisting mode, the ${}^{28}\rm {Si}$ nuclei rotate 
around the molecular $z'$-axis, and so the spin vectors are parallel to 
it. Since the fragments $^{28}\rm Si$ are detected approximately at 
$\theta_{\rm cm}=90^\circ$ (the (c)-axis direction), 
the molecular $z'$-axis and the spin vectors are expected to be 
approximately parallel to (c)-axis. However our theoretical prediction 
does not exhibit the "m=2" pattern in (c). Unexpectedly, it appears in 
(a)-axis (the beam axis), as is displayed in Fig.~9 with dashed lines, 
which gives rise to a puzzle. 
This is an interesting aspect of extremely-high spin resonances. 
Due to the high-speed rotational motion, the final velocity vectors 
of the fragments are considered to be almost perpendicular 
to the relative vector of the fragments in the decay process, 
like raindrop motions splashed from a rapidly rotating umbrella. 
In order to examine the motions of the ${}^{28}\rm {Si}$ nuclei 
after the decay, we have analyzed classical orbits and obtained 
results that for $J=38$ the relative vector turns round 
about $70^\circ$ from the initial angle. 
As a confirmation, we calculated angular correlations with a relatively 
lower angular momentum $J=8$. 
The results show the "m=2" dominance in (c)-axis, as expected 
by the above explanation.

\subsection{Strengths of confinements} 

With respect to the very narrow resonances observed as correlating 
among the elastic and inelastic scattering, partial decay widths 
are investigated.   %%% as well as the fragment angular distributions. 
In the butterfly and the twisting modes, the decay probability amplitudes 
concentrate to the elastic and mutual $2^+$ channels. 
This is a characteristic feature of the normal mode excitations, 
which is due to the symmetric motions of the two $^{28}\rm Si$ nuclei.
Such characteristics are, however, not seen in the data at 
$E_{\rm cm}=55.8$MeV. 
The molecular ground state or the radially-excited state obtained 
by using the folding potential shows rather large elastic widths 
and weak excitations to the inelastic channels. 
Hence, the properties of the partial widths observed at 
$E_{\rm cm}=55.8$MeV have not been well reproduced for all those states.

It is worth to have a close look at the %%%characteristic 
results with the folding-model potential. 
The molecular ground state configuration obtained by using the folding 
potential with the frozen density approximation, 
exhibits very weak confinement in the twisting degree of freedom, 
which results in an almost-free rotation.  
Intuitively, the extreme weak or shallow potential in the twisting motion 
appears unphysical, because the two $^{28}\rm Si$ nuclei are touching 
with each other. 
That is, the folding model with the frozen density is not well adequate. 
Actually, the modification of the confinement in terms of larger values of
$\hbar \omega$ for the twisting and consistently for the butterfly motions 
shows that the partial decay width in the elastic channel greatly reduces, 
while those in the inelastic channels relatively increase. 
As a result, the magnitudes of the partial widths become to be 
consistent with the experimental characteristics, 
which finally gives good agreements with the experimental data 
for the fragment angular distributions. 
Thus, the analyses suggest a substantial change of the deformation 
of the constituent nuclei and/or formation of a neck between them, 
which gives rise to a stronger confinement than that obtained 
by the folding model. 
This is exactly consistent with the existence of the stable triaxial 
configuration. 
In addition, experimental observations suggesting the stronger 
confinement are known in the $^{24}\rm Mg + {}^{24}Mg$ system. 
Only 30\% of the resonance flux appears in the binary decay channels 
of $^{24}\rm Mg + {}^{24}Mg$.\cite{Saini}  %%[6].
A search for the remaining flux to the fusion-evaporation 
channels has been recently made.\cite{Salsac}  %%[39]. 
Those observations suggest that two incident nuclei interact with each 
other strongly with their density overlapping and induced deformations. 
Their effects would be described effectively by a stronger confinement. 
A mechanism for the strong confinement is intriguing and yet to be 
clarified theoretically.

\subsection{Observation conditions} 

Finally, observation conditions of resonances in heavy-ion reactions 
should be discussed.
In experimental excitation functions of the $^{28}\rm Si +{}^{28}Si$ 
system which are angle-averaged around $\theta_{\rm cm}=90^\circ$, 
background yields of the elastic scattering and of the single and mutual 
$2^+$ excitations decrease rapidly as the bombarding energy increases 
toward the resonance region.\cite{BettsPRL2,SainiBetts}  %%[6,7] 
The same is the case in the $^{24}\rm Mg +{}^{24}Mg$ system.\cite{Zurm}%%[8] 
Such low background yields are one of the conditions to clearly 
observe resonances.
Especially in the elastic scatterings, 
strong damping of non-resonant amplitudes is a necessary condition,
because the resonance yields themselves are very small. 
Over those background yields, many prominent peaks would be observed
with relatively large yields in the channels concerned,
which theoretically have to be well described by the resonance terms, 
such as by the $R$-matrix theory. 
Strong absorption by the smooth cutoff model has worked as a 
successful description for the elastic background scattering in the 
angular distributions, as well as appropriate optical model analyses.
Actually, in the $^{28}\rm Si +{}^{28}Si$ system, it is shown that the 
effects of the strong absorption do not necessarily work to eliminate 
resonance structures, but show up very narrow and prominent structures 
over the very low background.
This is quite different from the conditions of the resonances 
observed in lighter systems such as  $^{12}\rm C +{}^{12}C$ 
and $^{16}\rm O +{}^{16}O$, where weak absorption 
plays a crucial role.\cite{BCMSuppl}

\section{Conclusions}

Resonances in the $^{28}\rm Si+{}^{28} Si$ system have been studied 
by means of the molecular model, in which the interacting dinuclear 
system is described by the molecular rotation and the internal 
collective variables for the orientations of the poles of the 
constituent nuclei. 
The molecular model predicts rotational spectra with a variety of 
intrinsic molecular states, such as the butterfly, twisting and etc.
In order to explore which one of the various molecular modes agrees 
with the experimental data on the resonance at $E_{\rm cm}=55.8$MeV, 
we have performed comprehensive analyses on available physical 
quantities, i.e., on partial widths, fragment angular distributions 
and fragment-fragment-$\gamma$ angular correlations.

Among the various modes, the molecular ground state with $J=38$ is 
successfully selected to be a candidate 
for the resonance observed at $E_{\rm cm}=55.8$ MeV.
We have found that for the resonance state, 
a nuclear molecule $^{28}\rm Si-{}^{28}Si$ of triaxial shape 
wobblingly rotates with a very high spin. 
The present results are the first discovery of the modes 
with $K$-mixing in heavy-ion resonance states, though the 
tilting or wobbling mode was once discussed in deep inelastic 
scattering processes.\cite{Randrup}

In the partial decay widths, each normal mode shows an interesting 
feature, i.e., resonance amplitudes appear to be enhanced 
in each relevant characteristic channels.
Moreover, the butterfly modes show fragment spins in the fragment 
direction ("m=2"), while the twisting mode does in the beam direction. 
Since each normal mode shows its own distinct characteristics of 
the angular correlations, 
it is possible to identify each excitation of the modes. 
Thus, angular correlation measurements are a powerful tool 
for the study of nuclear structures of heavy-ion resonances. 
Systematic measurements are desired, not only on the resonance 
at $E_{\rm cm}=55.8$MeV. For example, the same kind of measurements 
on the other nearby resonances of $^{28}\rm Si+{}^{28}Si$ is 
strongly called for.

\section*{Acknowledgements}
The authors thank Drs. C. Beck, R. Freeman and F. Haas 
for stimulating discussions in their collaborations.
The authors are grateful for the discussion and 
for the hospitality of Dr. B. Giraud in the visits at Saclay.

This work was supported in part 
by the Grant-in-Aid for Scientific Research from the Japanese 
Ministry of Education, 
Culture, Sports, Science and Technology (12640250).

\appendix

\section{%%Fifth Appendix
Relation between Molecular Wave Functions 
     and Channel Wave Functions for Calculating Decay Properties}

The total system consists of two deformed nuclei, for which 
we have assumed the axial symmetry and constant deformations. 
We have seven degrees of freedom illustrated in Fig.~12(a) 
except for the center of mass motion for the whole system, that is, 
the relative vector $\hbox{\mbf R} =(R,\theta_{2},\theta_{1})$ 
and the orientations of the deformations of the interacting nuclei 
described with Euler angles 
$({\tilde \alpha}_1, {\tilde \beta}_1)$ 
and $({\tilde \alpha}_2, {\tilde \beta}_2)$. 
Thus with those variables, generalized channel wave functions of 
$\{(I_1 I_2) I l , J M \}$ scheme are defined as usual 
in the $R$-matrix theory.
Specifying the spins $I_1$ and $I_2$ for the two constituent nuclei, 
a function with channel spin $I$ is given by
\begin{equation}
\psi_{(I_1 I_2) I M_I}= \sum_{M_1 M_2} (I_1 I_2 M_1 M_2| I M_I)
    \chi_{I_1 M_1}({\tilde \alpha_1}, {\tilde \beta_1})
    \chi_{I_2 M_2}({\tilde \alpha_2}, {\tilde \beta_2}),
\label{eq:A.1}
\end{equation}
where $\chi_{I_i M_i}({\tilde \alpha_i}, {\tilde \beta_i})$ 
denote internal wave functions for the constituent nuclei $^{28}\rm Si$. 
The generalized channel wave function is given 
with the orbital angular momentum $l$ for the relative motion, 
\begin{equation}
\hbox{\calg Y}_{(I_1 I_2) I l, J M} = \sum_{M_I, m} (I l M_I m | J M)
   \hbox{\calg S}_{12} 
\big[ \psi_{(I_1 I_2) I M_I} Y_{lm}(\theta_2, \theta_1) \big],
\label{eq:A.2}
\end{equation}
where 
$\hbox{\calg S}_{12}=
          (1+\hbox{\calg P}_{12})/{\sqrt {2(1+\delta_{I_1 I_2})}}$ 
denotes the symmetry operator between two $^{28}\rm Si$ nuclei.  
The internal wave functions in Eq.~(\ref{eq:A.1}) are assumed 
to be a rotational type, such as, 
\begin{equation}
\chi_{I_i M_i}(
{\tilde \alpha_i}, {\tilde \beta_i})=
\sqrt{2I_i +1 \over 8\pi^2} D^{I_i}_{M_i 0}(
{\tilde \alpha_i}, {\tilde \beta_i}, {\tilde \gamma_i}),
\label{eq:A.3}
\end{equation}
where $K_i=0$ are given in consistent with 
the axial symmetry of the constituent nuclei assumed, 
and hence ${\tilde \gamma_i}$ are spurious and do not appear 
in the l.h.s of Eq.~(\ref{eq:A.3}).

%%%%%%%%%%%%%%%%%%%%%%%%%%%%%%%%%%%%%%%%%%%%%%%%%%
%%%%%%%%%%%%%%%%%%%%%%%%%%%%%%%%%%%%%%%%%%%%%%%%%%
\begin{figure}[b] %%[htb]
 \parbox[b]{\halftext}{
   \begin{center}
     \includegraphics[scale=0.5]{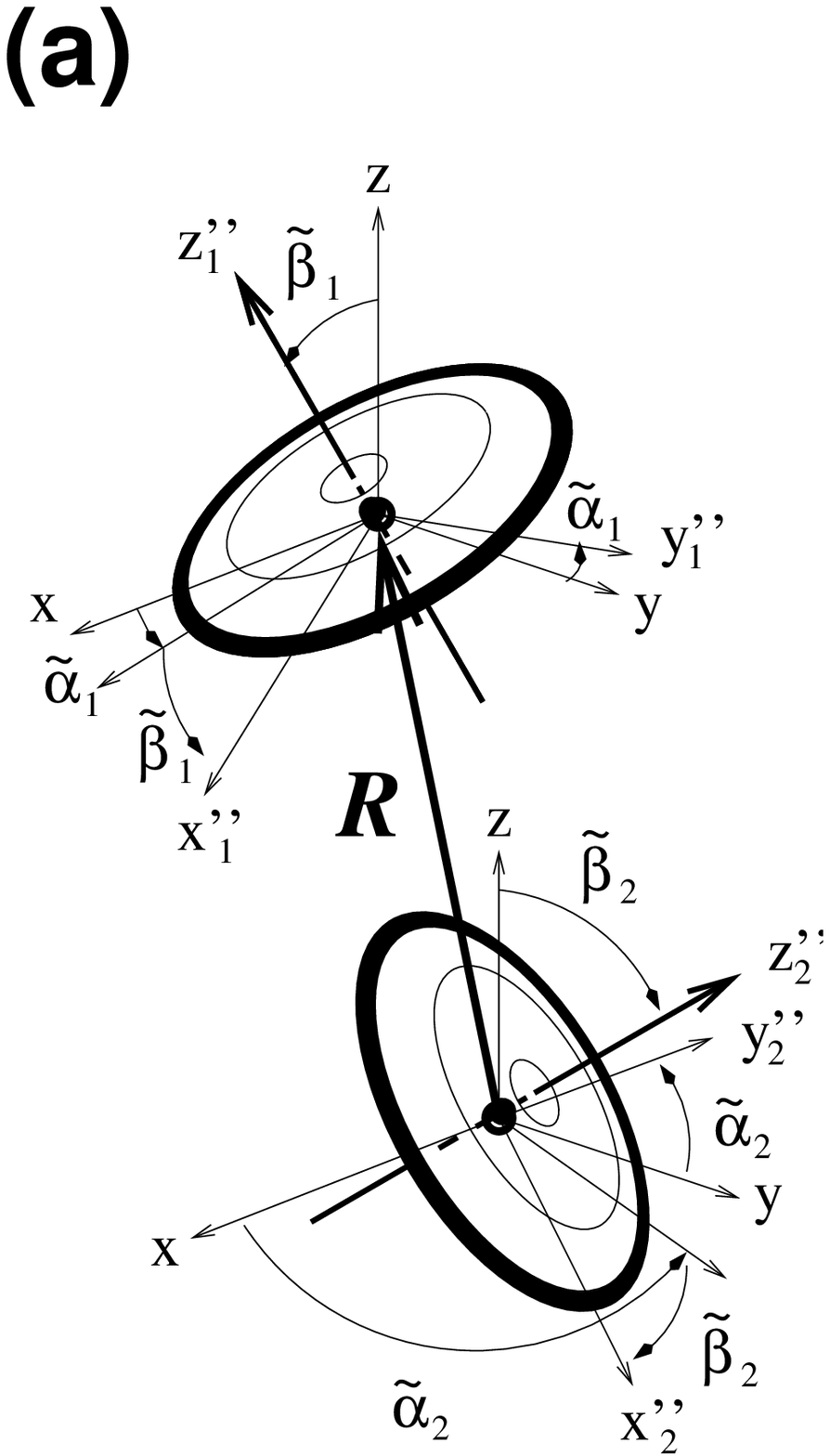}
   \end{center}
 }
 \parbox[b]{\halftext}{
   \begin{center}
     \includegraphics[scale=0.5]{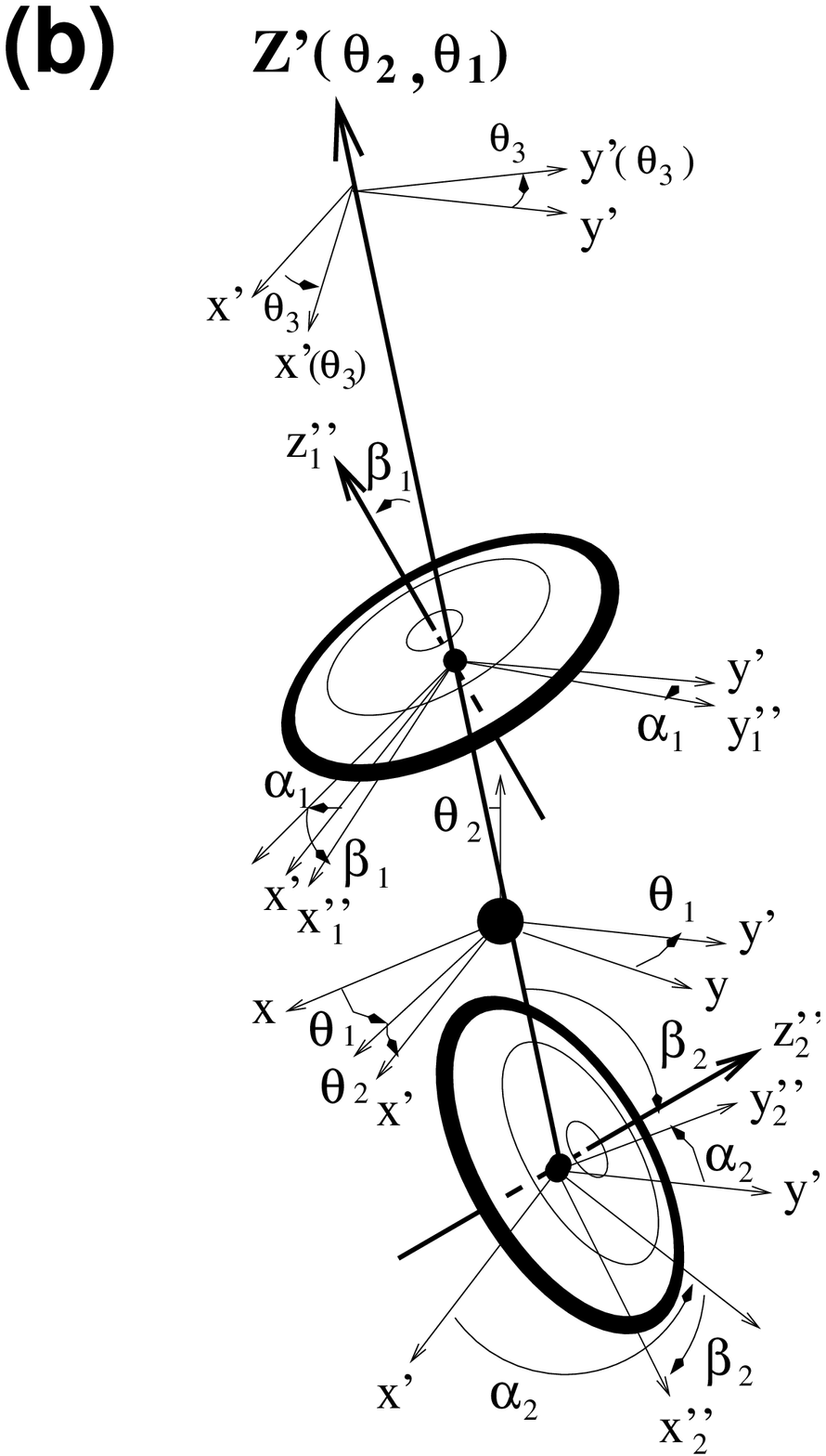}
   \end{center}
 }
\caption{The coordinates of an interacting dinuclear system. 
(a) shows the relative vector $\boldsymbol{ R} =(R,\theta_2,\theta_1)$ 
and usual Euler angles $({\tilde \alpha}_i, {\tilde \beta}_i)$ 
of the $i$-th nucleus referring to the laboratory frame. 
In (b), the molecular $z'$-axis and the seven degrees of freedom 
of the system are displayed, 
where the distance $R$ is not indicated explicitly. 
The third angle $\theta_3$ is defined by 
$\theta _{3} =(\alpha_1 + \alpha_2)/2$ 
to give the whole rotation around the $z'$-axis.
The figure is taken from Ref.~\citen{UeNewI}.
}
                 \label{fig:12}
%%%%%%%%%%%%%%%%%%%%%%%%%%%%%%%{Fig2a-labsys-obob.eps}PaperI
%%%%%%%%%%%%%%%%%%%%%%%%%%%%%%%{Fig2b-molsys-obob.eps}PaperI
\end{figure}
%%%%%%%%%%%%%%%%%%%%%%%%%%%%%%%%%%%%%%%%%%%%%%%%%%

Next we define variables in the molecular coordinate system 
for the model wave functions. 
As is illustrated in Fig.~12(b), 
we define a rotating molecular axis $z'$ of the whole system 
with the direction of the relative vector $\hbox{\mbf R}$ 
of the two interacting nuclei. 
In the molecular model, orientations and/or motions of 
the intrinsic axes of each deformed nucleus are 
referred to the molecular axes as usual.
For describing those orientations, we introduce new Euler angles 
$(\alpha_1,\beta_1,\gamma_1)$ and $(\alpha_2,\beta_2,\gamma_2)$, 
relations of which with the Euler rotations 
 $({\tilde \alpha_i}, {\tilde \beta_i}, {\tilde \gamma_i})$ 
referred to the laboratory system are given as
\begin{equation}
  \Omega_i ({\tilde \alpha}_i, {\tilde \beta}_i, {\tilde \gamma}_i)
        =   \Omega'_i (\alpha_i, \beta_i, \gamma_i) 
             \Omega_M (\theta_1, \theta_2) , \qquad   i=1,\,\,  2,
\label{eq:A.4}
\end{equation}
where $\Omega_i$ and $\Omega'_i$ denote Euler rotations 
for the $i$-th constituent nucleus with respective angles, and 
$\Omega_M$ denotes rotations of the molecular axes.
On the r.h.s. of Eq.~(\ref{eq:A.4}) $\Omega'_i$ denote 
the successive rotations after $\Omega_M$; 
firstly the axes of the $i$-th constituent nucleus rotate 
up to the directions of the molecular axes by $\Omega_M$, 
and secondly they rotate referring to the molecular axes by $\Omega'_i$.
Since the second rotations are written as 
$\Omega'_i (\alpha_i, \beta_i, \gamma_i) = \Omega_M (\theta_1, \theta_2) 
\Omega_i(\alpha_i, \beta_i, \gamma_i) \Omega^{-1}_M (\theta_1, \theta_2)$ 
with rotations $\Omega_i(\alpha_i, \beta_i, \gamma_i)$ referring 
to the laboratory system, we obtain 
\begin{equation}
\Omega_i ({\tilde \alpha}_i, {\tilde \beta}_i, {\tilde \gamma}_i)
     =\Omega_M (\theta_1, \theta_2) \Omega_i (\alpha_i, \beta_i, \gamma_i),
\label{eq:A.5}
\end{equation}
and accordingly we are able to define 
the new Euler angles $(\alpha_i,\beta_i,\gamma_i)$ by the relations 
\begin{equation}
 \Omega_i (\alpha_i, \beta_i, \gamma_i) 
        =   \Omega^{-1}_M (\theta_1, \theta_2) 
        \Omega_i ({\tilde \alpha}_i, {\tilde \beta}_i, {\tilde \gamma}_i).  
\label{eq:A.6}
\end{equation}
Note that $\gamma_i$ are spurious as well as ${\tilde \gamma}_i$, 
and so we neglect them in the following descriptions.

The molecular $x'$-axis would be determined with the two constituent 
nuclear configurations around the $z'$-axis which are specified 
by $\alpha_i$.
Hence they are combined 
into $\theta _{3} =(\alpha _{1} + \alpha _{2})/2$ and 
$\alpha  =(\alpha _{1} - \alpha _{2})/2$, 
and then we have molecular coordinates, 
\begin{equation}
 (q_{i})=
 (\theta _{1},\theta _{2},\theta _{3}, R, \alpha,\beta _{1},\beta _{2}),
\label{eq:A.7}
\end{equation}
where $\theta_1, \theta_2$ and $\theta_3$ 
are the Euler angles of the rotating molecular frame 
with the other fours being internal variables. 
The molecular $x'$- and $y'$-axes defined with $\theta_3$ are 
illustrated as $x'(\theta_3)$ and $y'(\theta_3)$ in Fig.~12(b).
With the definition of the molecular $x'$-axis with $\theta _3$, 
rotational angles of each constituent nucleus should be redefined 
so that 
$\alpha_1(\theta_3)=\alpha_1 - \theta_3 = \alpha$ 
and
$\alpha_2(\theta_3)=\alpha_2 - \theta_3 = -\alpha$.
Here after we write them as $\alpha_i$ simply, and 
correspondingly the relations Eq.~(\ref{eq:A.5}) are rewritten as 
\begin{equation}
 \Omega_i ({\tilde \alpha}_i, {\tilde \beta}_i)
        =   \Omega_M (\theta_1, \theta_2, \theta_3) 
             \Omega_i (\alpha_i, \beta_i).
\label{eq:A.8}
\end{equation}

Now, by means of relations (\ref{eq:A.8}), the internal wave functions 
of Eq.~(\ref{eq:A.3}) are written with the molecular coordinates as 
\begin{equation}
\chi_{I_i M_i}({\tilde \alpha_i}, {\tilde \beta_i}) 
  =\sum_{\mu_i}
         D^{I_i}_{M_i \mu_i}( \theta_1,\theta_2,\theta_3)
         \chi_{I_i \mu_i}( \alpha_i, \beta_i). 
\label{eq:A.9}
\end{equation}
We substitute Eq.~(\ref{eq:A.9}) into Eqs. (\ref{eq:A.1}) 
and (\ref{eq:A.2}), and use the following relations, 
\begin{equation}
\sum_{M_1 M_2} (I_1 I_2 M_1 M_2| I M_I)
D^{I_1}_{M_1 \mu_1}( \theta_i) D^{I_2}_{M_2 \mu_2}( \theta_i)
 =
(I_1 I_2 \mu_1 \mu_2 | I \, \mu_1+\mu_2) 
                  D^I_{M_I, \, \mu_1+\mu_2}( \theta_i),
\label{eq:A.10}
\end{equation}
\begin{eqnarray}
 \sum_{M_I, m} (I l M_I m | J M)  %%&
D^{I}_{M_I, \, \mu_1+\mu_2}( \theta_i) Y_{lm}( \theta_2,\theta_1)=
\nonumber \\
 (I l \, \mu_1+\mu_2 \,\, 0| J \,\, \mu_1+\mu_2) 
{\sqrt{(2l+1)/ 4\pi}} \, 
            D^J_{M, \, \mu_1+\mu_2}(\theta_i),
\label{eq:A.11}
\end{eqnarray}
where $\theta_i$ is the abbreviation for $\theta_1,\theta_2,\theta_3$.
Thus we obtain the generalized channel wave function expressed with 
the molecular coordinates,
\begin{equation}
\hbox{\calg Y}_{(I_1 I_2) I l ; J M} = 
 \sum_K (I l K 0 | J K)  
{\sqrt{2l+1 \over 4\pi}} 
   \hbox{\calg S}_{12} 
\big[ D^J_{M K}(\theta_i)
\Phi_{(I_1 I_2) I K}(\alpha, \beta_1, \beta_2) \big], 
\label{eq:A.12}
\end{equation}
with
\begin{eqnarray}
\Phi_{(I_1 I_2) I K}(\alpha, \beta_1, \beta_2) &= &
 \sum_{\mu_1 \mu_2} (I_1 I_2 \mu_1 \mu_2| I K)
    \chi_{I_1 \mu_1}( \alpha, \beta_1)
    \chi_{I_2 \mu_2}( -\alpha, \beta_2) 
\nonumber \\
&=& \sum_{\mu_1 \mu_2} (I_1 I_2 \mu_1 \mu_2| I K) \, {\rm e}^{i\nu \alpha}
  d^{I_1}_{ \mu_1 0}(\beta_1) d^{I_2}_{ \mu_2 0}(\beta_2), 
\label{eq:A.13}
\end{eqnarray}
where $\nu \equiv \mu_1 - \mu_2$ denotes twisting quantum number.
As the generalized channel wave functions are expanded with 
the rotational functions for the whole system 
in a series of $K=\mu_1+\mu_2$, 
we are able to easily calculate overlapping integrals between 
each channel wave function with $\{ (I_1 I_2) I l, J M \}$ 
and the model wave functions specified with $\{ J M K \}$.
Note that in the $R$-matrix formula in \S2   %%section 2 
and for the generalized channel wave functions given above, 
normalization integrals are defined as usual, 
while the definition of the model wave functions $\Psi_\lambda$ 
in Eqs.~(\ref{eq:2.27}) and (\ref{eq:2.28}) is given 
for the vibrational modes with the volume element 
$dV =dR d\alpha d\beta_1 d\beta_2$ 
(see \S2.2 of the paper~I\cite{UeNewI} for details). 
Hence, for the internal wave functions $\Psi^{JM}_\lambda$ 
in Eq.~(\ref{eq:2.5}), we use $\Psi_\lambda /\sqrt D$, 
where $\Psi_\lambda$ denote the model wave functions 
and $D= \mu^{3/2}R^2 I_1 \sin\beta_1 I_2 \sin\beta_2$, 
square root of which is the additional phase factor 
of the molecular model wave functions.

Finally we mention the property of the channel functions 
described in the molecular coordinate system.
Due to the angular momentum coupling 
 $(I_1 I_2 \mu_1 \mu_2| I K)$, we have restrictions 
$\mu_1 \le I_1$, $\mu_2 \le I_2$ and $K=\mu_1+\mu_2 \le I$ 
for the summations on $K$ and $(\mu_1, \mu_2)$ in 
Eqs.~(\ref{eq:A.12}) and (\ref{eq:A.13}).
For the elastic channel, where $I_1$ and $I_2$ are to be equal 
to $0$, the values of $\mu_1$ and $\mu_2$ are zero 
and so the same as for $K$ and $\nu$.
Hence the elastic channel wave function has no real 
$\alpha$-dependence nor $\beta_i$-dependence, and it is described 
simply by $D^J_{M 0}(\theta_1,\theta_2,\theta_3)$, 
i.e., by $Y_{JM}( \theta_2,\theta_1)$.
Thus, due to the boson symmetry, we have only positive-parity states 
with $l=J=even$, as usual.

\end{document}